\documentclass[journal=jacsat,manuscript=article]{achemso}
\setkeys{acs}{maxauthors=0}
\SectionNumbersOn 

\usepackage{amsfonts}
\usepackage{physics}
\usepackage{braket}
\usepackage{qcircuit}
\usepackage{graphicx}
\usepackage{dcolumn}
\usepackage{bm}
\usepackage[hidelinks]{hyperref}
\usepackage{tensor}
\usepackage{tabularx} 
\usepackage{booktabs}
\usepackage{multirow}
\usepackage{xurl}
\usepackage{xcolor}
\usepackage{chemformula}
\usepackage{threeparttable}
\usepackage[ruled, linesnumbered]{algorithm2e}

\newcommand{\ourmethodU}{U_{\textrm{\textbf{CHEM}}}}

\newcommand{\rev}[1]{\textcolor{black}{#1}}
\usepackage{qcircuit}

\title{Efficient and Robust Parameter Optimization of the Unitary Coupled-Cluster Ansatz}

\author{Weitang Li}
\email{liwt31@gmail.com}
\affiliation{Tencent Quantum Lab, Tencent, Shenzhen, 518057, China}

\author{Yufei Ge}
\affiliation{Department of Chemistry, Tsinghua University, Beijing, 100084, China}

\author{Shi-Xin Zhang}
\affiliation{Tencent Quantum Lab, Tencent, Shenzhen, 518057, China}

\author{Yu-Qin Chen}
\affiliation{Tencent Quantum Lab, Tencent, Shenzhen, 518057, China}

\author{Shengyu Zhang}
\affiliation{Tencent Quantum Lab, Tencent, Hong Kong, 999077, China}

\begin{document}

\begin{abstract}
The variational quantum eigensolver (VQE) framework has been instrumental in advancing near-term quantum algorithms. 
However, parameter optimization remains a significant bottleneck for VQE, requiring a large number of measurements for successful algorithm execution. 
In this paper, we propose sequential optimization with approximate parabola (SOAP) as an efficient and robust optimizer specifically designed for parameter optimization of the unitary coupled-cluster ansatz on quantum computers. 
SOAP leverages sequential optimization and approximates the energy landscape as quadratic functions, minimizing the number of energy evaluations required to optimize each parameter. 
To capture parameter correlations, SOAP incorporates the average direction from the previous iteration into the optimization direction set. Numerical benchmark studies on molecular systems demonstrate that SOAP achieves significantly faster convergence and greater robustness to noise compared to traditional optimization methods. Furthermore, numerical simulations up to 20 qubits reveal that SOAP scales well with the number of parameters in the ansatz. The exceptional performance of SOAP is further validated through experiments on a superconducting quantum computer using a 2-qubit model system.

\end{abstract}

\maketitle

\section{Introduction}
Chemistry has emerged as a promising field for the application of noisy intermediate-scale quantum (NISQ) devices~\cite{bharti2022noisy, hoefler2023disentangling, mcardle2020quantum, ma2023multiscale}.
The key component for the success is the variational quantum eigensolver (VQE) framework, which leverages quantum computers as specialized devices for storing the wave function of molecular systems~\cite{peruzzo2014variational, cerezo2021variational, tilly2022variational}.
More specifically, a quantum circuit, characterized by a parameterized unitary transformation $\hat U(\vec \theta)$ and the initial state $\ket{\phi_0}$, is viewed as an ansatz $\ket{\phi(\vec \theta)} = \hat U(\vec \theta)\ket{\phi_0}$ analog to the ansatz in classical computational chemistry. 
Leveraging the Rayleigh-Ritz variational principle,
the parameters in the quantum circuit $\vec \theta$ are optimized on a classical computer using either gradient-based or gradient-free optimizer,
with the energy expectation $E(\vec \theta)=\braket{\phi|\hat H |\phi}$ measured on quantum computers.

Two distinct types of ansatz have been explored in quantum computational chemistry.
The disentangled unitary coupled-cluster (UCC) ansatz~\cite{anand2022quantum} 
uses the Hartree-Fock (HF) state as the initial state, and applies the UCC factors $e^{\theta_j \hat G_j}$ to the initial state,
where $\vec \theta$ are circuit parameters and $\{\hat G_j\}$ are anti-Hermitian cluster operators.
$\hat G_j$ is composed of excitation operators and it is mostly commonly truncated to single and double excitation.
The UCC ansatz is the most widely studied ansatz due to its accuracy and close relation with classical computational chemistry, yet its circuit depth is usually too deep to be implemented on currently available quantum devices~\cite{lee2018generalized, evangelista2019exact}.
\rev{
To address this issue, several variants of the UCC ansatz have been proposed. 
For example, adaptive derivative-assembled pseudo-Trotter ansatz VQE (ADAPT-VQE)~\cite{grimsley2019adaptive} and paired UCC with double excitations~\cite{elfving2021simulating}  aim to reduce the circuit depth while preserving the clear chemical structure of the UCC ansatz.
}
On the other hand, the hardware-efficient ansatz (HEA) features shallow circuit depth and finds widespread application in hardware experiments~\cite{kandala2017hardware}. However, its scalability is hindered by optimization challenges, which need to be addressed for broader practical application~\cite{mcclean2018barren}.
\rev{
Furthermore, 
there is a growing interest in integrating the hardware-efficient ansatz with chemical and physical insights~\cite{sun2024toward, xiao2024physics}.
}

One of the biggest bottlenecks for VQE is the excessive number of measurements required to achieve high accuracy.
Various basis rotation methods have been proposed to mitigate this issue~\cite{verteletskyi2020measurement, huggins2021efficient, wu2023_Z}. 
Meanwhile, the pursuit of an efficient parameter optimization subroutine has been a central focus in recent research~\cite{rivera-dean_avoiding_local_minima_z, PRR_local_minima_z, tamiya2022stochastic, cheng2023error, choy2023molecular, Liu2023a_z}.
Parameter optimization for quantum circuits presents notable challenges due to the large number of parameters involved and the necessity to find the global minimum. Moreover, the efficient computation of energy gradients on a quantum computer remains unfeasible.
Specifically,  computing all the gradients requires $O(N)$ circuit evaluations, where $N$ is the number of parameters, \rev{using the parameter-shift rule}~\cite{mitarai2018quantum, schuld2019evaluating}.
In contrast, classical computers can leverage the back-propagation algorithm to compute all gradients with a time complexity of 
$O(1)$~\cite{rumelhart1986learning}.
\rev{We note that the time complexity is relative to 
the computation of the target function.}
In other words, computing all gradients requires roughly the
same time as evaluating the function itself.
\rev{
Recently, a quantum gradient algorithm with $O(1)$ scaling has also been reported~\cite{bowles2023backpropagation}.
However, the algorithm imposes restrictions on the circuit structure and its applicability to chemical problems remains to be explored.
}
Another fundamental distinction between quantum computers and classical computers lies in the inherent uncertainty of measurement outcomes in quantum systems, emphasizing the need for noise-resilient optimizers.
The difficulty associated with parameter optimization is manifested by the growing popularity of pre-optimizing the circuit parameters on a classical simulator in recent hardware experiments~\cite{kawashima2021optimizing, o2023purification, zhao2023orbital}.

Unfortunately, there has been relatively limited research dedicated to the development of efficient parameter optimization algorithms for quantum computational chemistry.
In most quantum computational chemistry research,
the prevailing practice involves employing general-purpose optimizers implemented in standard packages such as SciPy~\cite{2020SciPy-NMeth}. 
For instance, the Limited-memory Broyden–Fletcher–Goldfarb–Shanno with bounds (L-BFGS-B) algorithm~\cite{byrd1995limited} demonstrates efficiency in numerical simulations where the gradients are readily available~\cite{lang2020unitary, liu2020simulating, khamoshi2022agp}.
Conversely, the constrained optimization by linear approximation (COBYLA) algorithm~\cite{powell1994direct} and the simultaneous perturbation gradient approximation (SPSA) algorithm~\cite{spall1992multivariate} are popular choices for hardware experiments as they do not rely on gradient information~\cite{kandala2019error, mccaskey2019quantum, eddins2022doubling}.
However, gradient-free optimizers often suffer from slow convergence rates, further exacerbating the already high number of measurements required for VQE. 
These optimizers are inefficient because they fail to fully exploit the unique mathematical structure of unitary quantum circuits and are typically sensitive to the measurement uncertainty inherent in quantum computers.
Consequently, there is an urgent need to develop efficient, gradient-free, and noise-resilient optimization algorithms specifically tailored to the unique characteristics of quantum computers and the context of computational chemistry.

Recently, a promising optimizer called NFT  or rotosolve optimizer~\cite{parrish2019jacobi, nakanishi2020sequential, ostaszewski2021structure} has emerged,  specifically designed for HEA circuits.
This method employs a parameter optimization approach where one parameter is optimized at a time while keeping the other parameters fixed, resembling the basic iteration procedure in Powell's method~\cite{powell1964efficient, press2007numerical, brent2013algorithms}.
By exploiting the fact that the energy expectation as a function of a single parameter can be expressed using a sine function, the NFT optimizer constructs an analytical expression for the energy expectation.
This construction only requires three independent energy evaluations, eliminating the need for computing gradients. 
The parameter is then directly optimized to the minimum, and the whole process can be considered as a variant of the usual line search procedures.
One notable advantage of the NFT optimizer is its exceptional tolerance to statistical measurement noise. This is achieved by avoiding comparisons of similar energy values, even when the circuit parameters are close to the minimum. 
However, a limitation of this method is that it is most efficient when each parameter is associated with a single-qubit rotation gate $\hat R_j(\theta_j) = e^{-i\frac{\theta_j}{2} \hat A_j}$ where $\hat A_j \in \{\hat X, \hat Y, \hat Z, \hat I\}$.
Thus, the NFT optimizer cannot be readily applied to the UCC ansatz without introducing significant overhead~\cite{kottmann2021feasible}.
In the UCC ansatz,
a UCC factor $e^{\theta_j\hat G_j}$  will be compiled into a quantum circuit in which multiple single qubit rotation gates share the same parameter $\theta_j$.
The number of parameter-sharing gates is $2^{2K-1}$, if the order of the excitation operator is $K$. 
Furthermore, the sequential optimization approach employed by the NFT optimizer neglects the ``correlation'' between the parameters.
In cases where the Hessian matrix $\pdv{E}{\theta_i \theta_j}$ has large non-diagonal elements,
the line search procedure only takes small steps towards the optimal parameter vector $\vec \theta$.
This can result in a large number of steps required for convergence, as illustrated in Fig.~\ref{fig:diagram}.
Consequently, sequential optimization algorithms may face challenges in converging efficiently for complex chemical systems~\cite{singh2023benchmarking}.

In this paper, we address these challenges with a parameter optimization algorithm tailored for the UCC ansatz.
Our method, termed sequential optimization with approximate parabola (SOAP), shares similarities with the NFT optimizer as it employs a sequential line-search procedure over a list of directions.
For each direction, SOAP obtains the minimum by fitting an approximate parabola using 2 to 4 energy evaluations. 
The key insight enabling this simple parabola fit is that for the UCC ansatz the initial guess is already in proximity to the minimum.
More specifically, the initial guess is usually generated by second-order Møller–Plesset perturbation theory~\cite{Romero19}, one of the most useful computational methods beyond the HF approximation.
By sacrificing the exact fit of the complicated energy function for each parameter, SOAP is also capable of performing a line search over the average direction of the last iteration. This feature significantly accelerates the convergence when there are large off-diagonal elements of the Hessian matrix.

To evaluate the performance of SOAP, we conduct extensive benchmark studies on systems ranging up to 20 qubits.
The results demonstrate that SOAP exhibits superior efficiency and noise resilience compared to traditional optimization methods.
Furthermore, our numerical data indicates that SOAP scales well with the number of parameters in the circuit.
For a UCC circuit with $N$ parameters, on average SOAP reaches convergences by $3N$ energy evaluations.
Our findings are verified on a superconducting quantum computer based on a simplified 2-electron, 2-orbital model.

\section{Methodology}
\subsection{Review and motivation}
\label{sec:rev-mot}
Let us begin by reviewing the NFT method.
Consider a quantum circuit parameterized by $\vec \theta$,
where for any parameter $\theta_j$, the circuit can be expressed as
\begin{equation}
    \ket{\phi(\vec \theta)} = \hat W_j \hat R_j(\theta_j) \hat V_j \ket{\phi_0} \ .
\end{equation}
Here, $\hat W_j$ represents the part of the circuit that depends on the parameters $\{\theta_k|k<j\}$,
and $\hat V_j$ represents the part that depends on $\{\theta_k|k>j\}$.
The rotation gates can be written as
\begin{equation}
\label{eq:rot_expansion}
    \hat R_j(\theta_j) = e^{-i\theta_j \hat A_j} = \cos\theta_j - i \hat A_j \sin\theta_j \ .
\end{equation}
Here we intentionally absorbed the $\frac{1}{2}$ factor into $\theta_j$ for notation simplicity.
It follows that the energy expectation as a function of a single parameter $\theta_j$ can be expressed as:
\begin{equation}
\label{eq:nft_fit}
\begin{aligned}
    E(\theta_j) & = a\cos^2{\theta_j} + 2b\sin\theta_j\cos\theta_j + c \sin^2\theta_j \\
    & = \sqrt{b^2 + \left(\frac{a-c}{2}\right)^2} \sin(\rev{2}\theta_j + \arctan{\frac{a-c}{2b}}) + \frac{a+c}{2} \,
\end{aligned}
\end{equation}
where $a$, $b$ and $c$ are some coefficients that depend on $\hat W_j$ and $\hat V_j$.
The minimum of $E(\theta_j)$ locates at $\theta_j = \rev{\frac{3}{4}}\pi - \rev{\frac{1}{2}}\arctan{\frac{a-c}{2b}}$.
By performing three independent energy evaluations at different $\theta_j$ values, the coefficients $a$, $b$ and $c$ can be determined.
Subsequently, the parameter $\theta_j$ is optimized to its optimal value while keeping the other parameters fixed.
The NFT method then proceeds to optimize the next parameter until certain convergence criteria are met.

The NFT optimization method described above can be extended to the UCC ansatz by expressing the UCC factors as~\cite{chen2021quantum, rubin2021fermionic}
\begin{equation}
\label{eq:expansion}
    e^{\theta_j G_j} = (1-\cos\theta_j)\hat G_j^2+\sin\theta_j \hat G_j + 1 \ .
\end{equation}
The energy expectation as a function of $\theta_j$ can then be written as
\begin{equation}
\label{eq:ucc_fit}
\begin{aligned}
    E(\theta_j) 
    = a_1 \cos^2\theta_j + a_2\sin^2\theta_j + a_3\sin\theta_j   \cos\theta_j + a_4 \sin\theta_j + a_5 \cos\theta_j\ ,
\end{aligned}
\end{equation}
where $\{a_i|1\le i \le 5\}$ is the set of coefficients to be determined.
Note that the constant term \rev{$a_6$ has been} merged into $a_1 \cos^2\theta_j + a_2\sin^2\theta_j$ \rev{by $a_6=a_6\cos^2\theta_j + a_6\sin^2\theta_j$}.
Obtaining these coefficients requires 5 independent energy evaluations at different $\theta_j$ values.
Thus, it seems the NFT optimizer can be migrated to UCC ansatz straightforwardly.
However, in closed-shell systems, which are commonly targeted by the UCC ansatz, UCC factors with complementary spin components are expected to share the same parameter~\cite{tsuchimochi2020spin, Mehendale23}.
In such cases, the number of energy evaluations required for an exact fit of the energy landscape increases to 14.
\rev{
This is because the analytic expression for $E$ with respect to a single variable $\theta_j$ has more parameters to be determined from measurements if two UCC factors share the same parameter $\theta_j$.
}
Moreover, due to the increased complexity of Eq.~\ref{eq:ucc_fit} compared to Eq.~\ref{eq:nft_fit}, Eq.~\ref{eq:ucc_fit} is more prone to overfitting \rev{due to the presence of statistical measurement noise and the limited number of data points}.
Therefore, it is desirable to develop a more robust and efficient scheme for sequential optimization of the UCC ansatz.
It is worth noting that the more complex expansion in Eq.~\ref{eq:expansion} and the parameter-sharing scheme
also complicate the evaluation of the energy gradients in the UCC ansatz, which prompts the development of gradient-free optimizers.

The SOAP method is based on the intuition that the initial guess for the UCC ansatz is very close to a highly favorable local minimum, if not the global minimum.
The most straightforward approach to initialize the UCC ansatz is to set $\vec \theta = 0$, resulting in $\ket{\phi(\vec \theta)}$ being the Hartree-Fock state.
A more efficient approach for parameter initialization is to use second-order Møller-Plesset perturbation theory (MP2) to generate the initial guess~\cite{Romero19}.
If $\theta_{ij}^{ab}$ is associated with the excitation from orbital $ij$ to orbital $ab$, it is initialized as
\begin{equation}
    \theta_{ij}^{ab} = \frac{h_{ijba}-h_{ijab}}{\varepsilon_i+\varepsilon_j-\varepsilon_a-\varepsilon_b} \,
\end{equation}
where $h_{pqrs}$ represents the two-electron integral and $\varepsilon_p$ is the HF orbital energy.
The amplitudes of single excitations are set to zero.
In the weak correlation regime, the difference between the full configuration interaction (FCI) energy $E_{\textrm{FCI}}$ and the HF energy $E_{\textrm{HF}}$, known as the correlation energy $E_{\textrm{corr}}$, is much smaller than the energy scale of the Hamiltonian.
In other words, if the quantum circuit is initialized randomly, the resulting energy would be much higher than $E_{\textrm{HF}}$.
From a wavefunction perspective, the HF state is by definition the most dominating configuration of the ground state wavefunction in weakly correlated systems, with MP2 serving as a good correction to the HF state.
In the strong correlation regime, the validity of the MP2 initial guess is an open question.
Nevertheless, as we shall observe from the numerical experiments, the MP2 initialization strategy proves to be effective in the strong correlation regime.
It is important to note that applying the  UCC ansatz to strongly correlated systems is not advocated in the first place,
due to its single-reference nature.
For strongly correlated systems, UCC in combination with CASSCF is preferred~\cite{sokolov2020quantum, de2023complete}.
Compared with traditional UCC, UCC-CASSCF improves the overlap between the initial state and the desired ground state, which justifies the application of SOAP to perform parameter optimization.
\rev{
We note that there are also other ways to do parameter initialization~\cite{filip2022reducing}, and we believe that the optimization problem for VQE could be solved through the synergistic development of both initialization and optimization methods.
}

If we assume $\vec \theta$ is sufficiently close to the minimum,
the energy function can be approximated by a parabola at arbitrary direction $\vec v$
\begin{equation}
\label{eq:para-fit}
    E(\vec \theta + x \vec v) \approx ax^2 + bx + c
\end{equation}
The coefficients $a$, $b$ and $c$ can be determined by 3 energy evaluations and then an offset $-\frac{b}{2a}\vec v$ is added to $\vec \theta$ to move it to the minimum along the direction $\vec v$.
By fitting a parabola we have reduced the number of measurements required and make the algorithm more robust to statistical noise.
Such an approximation also allows more flexibility on how to choose $\vec v$, which will be discussed in the next section.

\subsection{The SOAP algorithm}
We first describe the basic iteration procedure for SOAP before delving into the treatment of corner cases and minor optimizations within SOAP.
The complete algorithm is presented in Algorithm~\ref{algo:soap}.
\rev{
We first initialize the direction vector $\vec v_i = \vec e_i$, where $\vec e_i$ is a unit vector in the parameter space.
In each iteration, for each direction vector $\vec v_i$,  the energies along $\vec \theta_{i-1} + x \vec v_i$  are measured. 
}
Here $\vec \theta_{i-1}$ is the optimized parameter vector from the previous optimization step along direction $\vec v_{i-1}$.
A parabola fit based on Eq.~\ref{eq:para-fit} is then performed using the measured energies.
Subsequently, the circuit parameters are optimized by setting $\vec \theta_i$ to  the minimum of the parabola, given by $\vec\theta_{i-1} -\frac{b}{2a}\vec v_i$. 
This process is in essence a simplified line search procedure over the direction $\vec v_i$ based on heuristics of the UCC ansatz.
The algorithm then proceeds to the next direction vector $\vec v_{i+1}$.
Once the entire list of direction vectors $\mathcal{V}=[\vec v_1, \vec v_2, \cdots, \vec v_N]$ has been traversed,
the SOAP algorithm then proceeds to the next iteration until certain convergence criteria are satisfied.
The initial order of $\mathcal{V}$ is determined by ensuring $|\theta_i| > |\theta_{i+1}|$ where
$\theta_i$ is $i$th element of the initial $\vec \theta$ determined by the MP2 amplitudes.

Let us now delve into more details about the line search procedure using an approximate parabola \rev{based on Eq.~\ref{eq:para-fit}}, denoted as LSAP in the following discussion.
To begin, we define a small scalar $u$ and let
\begin{equation}
    x_{m} = mu, m \in \mathbb{Z} \ .
\end{equation}
For example, $x_0 = 0, x_1 = u$.
We then define
\begin{equation}
 y_{m} = E(\vec \theta_{i-1} + x_m \vec v_i)   \ ,
\end{equation}
which is measured on a quantum computer.
Since we assume $\vec \theta_{i-1}$ is close to the minimum, we use the set $\mathcal{F} = \{(x_i, y_i)|i=-1, 0, 1\}$ to construct the parabola.
Note that only $y_{-1}$ and $y_1$ should be measured on a quantum computer since $y_0$ is available from the previous iteration.

Next, we turn our attention to the selection of $u$. Based on the landscape shown in Fig.~\ref{fig:landscape_n2}, setting $u=0.1$ is a reasonable initial value.
\rev{In the Supporting Information, we have included benchmark data that demonstrates that the performance of SOAP is not significantly affected by the choice of $u$ within the range of 0.05 to 0.20.}
Although it may be tempting to dynamically adjust $u$ according to the landscape during the optimization, we refrain from such enhancements for the following reasons.
Firstly, quantum computers are subject to measurement uncertainty, and shrinking $u$ will decrease the signal-to-noise ratio.
As a result, shrinking $u$ around the minimum does not necessarily lead to a more accurate estimation of $\vec \theta_i$.
Secondly, enlarging $u$ does not offer significant potential either, since the landscape of $E(\vec \theta_{i-1} + x\vec v_i)$ is periodic with a period of $2\pi$ and a good parabola fit can only be expected within a span of $\sim \pi/2$.
Lastly, we prefer keeping SOAP a simple and robust algorithm that is straightforward to implement and extend.
For these reasons, we use $u=0.1$ throughout the paper.

It is worth mentioning that the idea of parabola fitting has also been proposed in a recent work~\cite{armaos2022efficient}. 
\rev{In each iteration, their method sets the optimization direction to the direction of the gradients and then employs the parabola fitting to determine the minimum.
Therefore, their method is gradient-based while our method is gradient-free.}
The Newton's method for optimization can also be considered as a second-order approximation of the optimization landscape,
in which the Taylor expansion is constructed based on local first and second-order derivatives.
In contrast, by setting $u=0.1$, the LSAP procedure differentiates from Newton's method by exploiting \rev{non-local} landscape information and thereby becomes more robust to measurement noise.

In the LSAP procedure, certain corner cases deserve careful consideration.
Because the fitting relies on only three data points, the fitted energy landscape can be severely distorted if $x_0$ is significantly distant from the minimum.
Although this scenario is relatively rare, it is inevitable for strongly correlated systems, as shown in Fig.~\ref{fig:landscape_n2}. 
The quality of the parabola fit can be roughly estimated by checking if $y_0 < y_{-1}$ and $y_0 < y_1$.
In such instances, the local minimum is bracketed by $x_{-1}$ and $x_1$.
Conversely, if the minimum of $\{y_i|i=-1, 0, 1\}$ is not $y_0$,  then the assumption of a parabolic landscape is likely invalid. 
Under such circumstances, it is possible to fit a parabola with $a<0$ which would direct $\vec \theta$ towards the maximum.
Thus, more data is required to accurately construct the parabola around the minimum.

In the following discussion, we assume the minimum is at $x_{1}$. For cases where the minimum is at $x_{-1}$, the treatment follows similarly.
If the minimum is at $x_{1}$, we proceed to explore $x_{4}$ and add $(x_4, y_4)$ to $\mathcal{F}$.
The parabola is then fitted using the least squares method
\begin{equation}
    \min{\sum_{(x_i, y_i)\in \mathcal{F}}(ax^2_i+bx_i+c-y_i)^2} \ .
\end{equation}
We do not expect that using slightly different data points such as $x_{3}$ or $x_{5}$ will have a significant impact on the overall performance of the method.
\rev{The arbitrary choice of the additional data point is based on the understanding that these corner cases are rare during the optimization process. Thus, our primary objective for these cases is to prevent significant errors, rather than choosing the best possible data point that maximizes the performance.}
In even rarer corner cases where the newly measured $y_{4}$ is smaller than $y_1$,
we directly use $(x_{4}, y_{4})$ as the minimum.
This avoids the possibility of fitting a parabola with $a>0$ while maintaining the simplicity of the algorithm.

To summarize, in typical cases where $x_0$ is close to the minimum, two energy evaluations are required to execute the line search.
In the worst-case scenario where $x_0$ is distant from the minimum, three energy evaluations are required.
A keen reader might point out that since we have assumed $y_0=E(\vec \theta_{i-1})$ is available from the previous iteration, another energy evaluation at $\vec \theta_i$ is required before completing the current iteration.
However, to avoid extra measurements, a shortcut can be taken. 
When the quality of the fitting is good and the minimum is bracketed by $x_{-1}$ and $x_1$,
we assign the minimum value of the fitted parabola,  $c-\frac{b^2}{4a}$, to the energy $E(\vec \theta_i)$.
We note that using $c-\frac{b^2}{4a}$ directly could potentially be more accurate than measuring the energy on quantum computers
considering the inherent uncertainty of measurement.
On the other hand, if the parabola is fitted with the input from $(x_{-4}, y_{-4})$ or $(x_4, y_4)$, an additional energy evaluation at $\vec \theta_i$ is performed, to avoid the bias introduced by the approximate parabola.
\rev{
In practice, we find that this approximation saves approximately one-third of the energy evaluations without compromising the accuracy of the energy estimation.
}
Taking this into account, the number of energy evaluations at each iteration ranges from 2 to 4 depending on the quality of the parabolic fitting.
The complete LSAP algorithm is presented in Algorithm~\ref{algo:line_search}.

\begin{algorithm}
\DontPrintSemicolon
\caption{Line Search with Approximate Parabola (LSAP)}
\label{algo:line_search}
\SetKw{Break}{break}
\SetKwFunction{MIN}{MIN}
\SetKwFunction{MeasureEnergy}{MeasureEnergy}
\SetKwInOut{Input}{input}\SetKwInOut{Output}{output}

\Input{Direction vector $\vec v_i$, circuit parameter $\vec \theta_{i-1}$, and the corresponding energy $E_{i-1}$}
\Output{Circuit parameter at the minimum $\vec \theta_i$ and the corresponding energy $E_i$}
\BlankLine
$y_{-1}$ = \MeasureEnergy{$\vec \theta + x_{-1} \vec v_i $} \;
$y_0 = E_{i-1}$ \;
$y_{1}$ = \MeasureEnergy{$\vec \theta + x_{1} \vec v_i $} \;
$y_m$ = \MIN{$y_{-1}, y_0, y_1$} \;
$\mathcal{F} = \{(x_j, y_j)|j=-1, 0, 1\}$ \;
\tcc{treat corner cases}
\If{$y_m$ is not $y_{0}$}{
    \lIf{$y_m$ is $y_{-1}$}{$k=-4$}\lElse{$k=4$}
    $y_k$ = \MeasureEnergy{$\vec \theta + x_k \vec v_i $} \;
    \eIf{$y_m <y_k$}{
    add $(x_k, y_k)$ to $\mathcal{F}$ \;
    }{
    \Return $\vec \theta + x_k \vec v_i$, $y_k$ \;
    }
}
Fit $y=ax^2+bx +c$ based on $\mathcal{F}$ \;
$\vec \theta_i=\vec \theta_{i-1} - \frac{b}{2a} \vec v_i$ \;
\lIf{$\mathcal{F}$ has exactly 3 elements}{$E_{i}=c-\frac{b^2}{4a}$}\lElse{$E_i$ = \MeasureEnergy{$\vec \theta_i$}}
\Return $\vec \theta_i$,  $E_i$\;
\end{algorithm}

\begin{figure}[h]
    \centering    \includegraphics[width=0.4\textwidth]{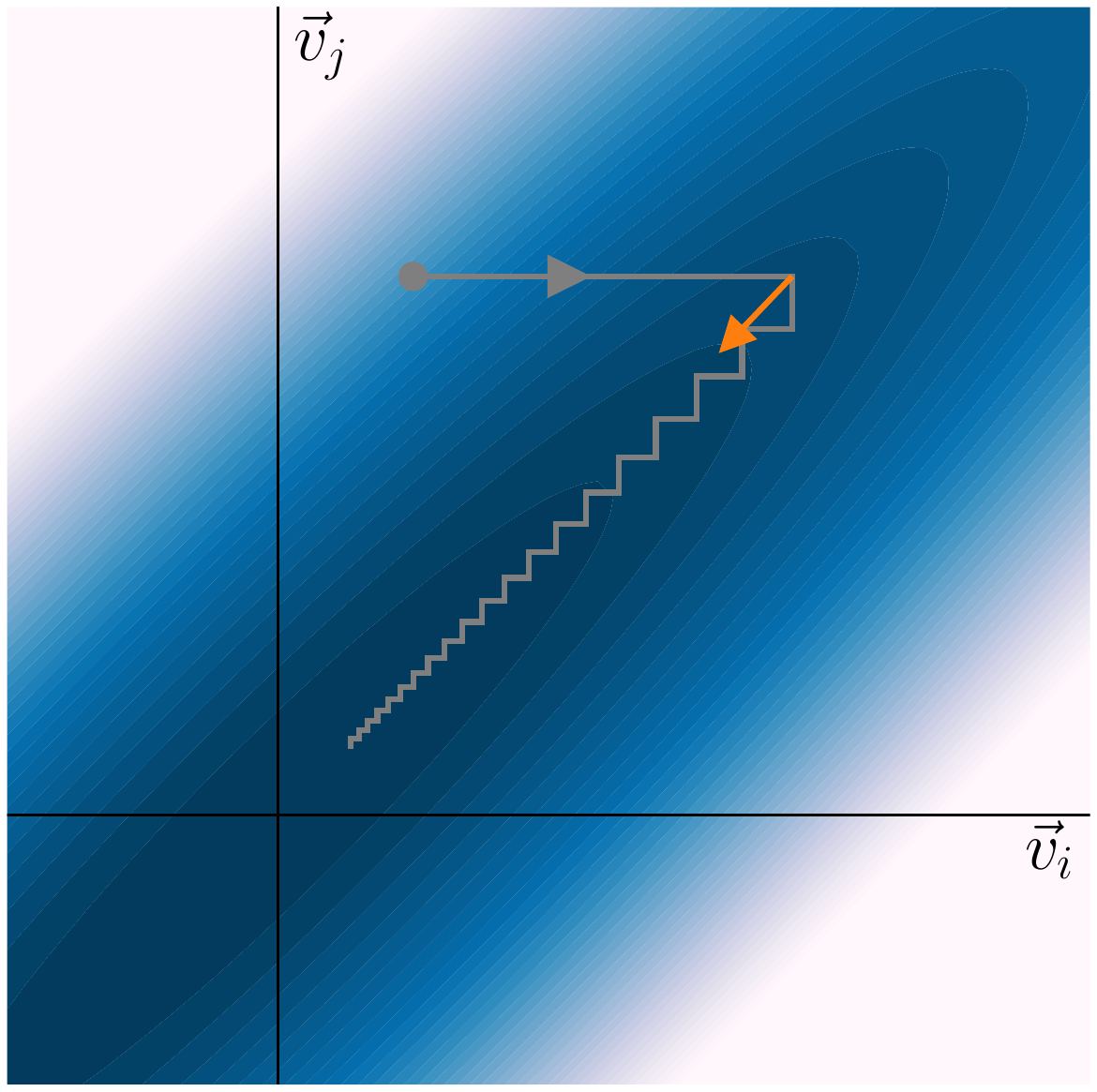}
    \caption{A schematic diagram showing the optimization trajectory of sequential optimization methods when there are large off-diagonal elements of the Hessian matrix. Traditional sequential optimization methods start from the gray dot and proceed along the gray line. SOAP identifies a more promising average direction marked by the orange arrow which significantly accelerates the convergence.}
    \label{fig:diagram}
\end{figure}

The sequential optimization method is most effective when the circuit parameters are ``uncorrelated'',
meaning that adjusting one of the parameters $\theta_i$ only causes a constant shift in the landscape with respect to other parameters $\theta_j$.
In numerical analysis, this property is referred to as $\vec v_i$ being ``conjugate'' with other $\vec v_j$.
However, one of the main limitations of the sequential optimization procedure is that it neglects the ``correlation'' between the circuit parameters.
More specifically, if the Hessian matrix $\pdv{\braket{E}}{\theta_i \theta_j}$ has large non-diagonal elements,
the optimization may take very small steps along $\vec v_i$ and $\vec v_j$ and results in a large number of
energy evaluations.
An example based on a function of two variables is depicted in Fig.~\ref{fig:diagram}.

To deal with this problem, we can draw inspiration from Powell's method.
Powell's method is a well-established method for multi-variable optimization which features iterative line-search over each of the variables~\cite{powell1964efficient}.
Powell's method updates the set of directions $\mathcal{V}$ to account for the missing ``correlation''.
After an iteration over $\mathcal{V}$,
denoting the set of parameters before and after as $\vec \theta_0$ and $\vec \theta_N$ respectively,
Powell's method adds $\vec \theta_N - \vec \theta_0$ to $\mathcal{V}$,
which represents the average optimization direction for the last iteration.
As shown in Fig.~\ref{fig:diagram}, this approach works well when the Hessian matrix has large non-diagonal elements.
Since $\mathcal{V}$ is now over-complete, an element from $\mathcal{V}$ can be dropped.
Powell's method suggests dropping $\vec v_j$, the direction with the largest energy descent $\Delta_j=|E_{j} - E_{j-1}|$,
where $E_j$ represents the estimated energy after optimization along $\vec v_j$.
This seemingly counter-intuitive choice is made because
$\vec v_j$ is likely to be a major component of the new direction $\vec \theta_N - \vec \theta_0$,
and dropping $\vec v_j$ avoids the buildup of linear dependence in $\mathcal{V}$.
In the context of SOAP, the new direction is further normalized to unit length since fixed $u$ is employed.

Powell's method also suggests several conditions when $\mathcal{V}$ should not be updated.
To assess the quality of the new direction $\vec \theta_N - \vec \theta_0$, an energy evaluation at an extrapolated point along the new direction $\vec \theta_N - \vec \theta_0$ is performed to get $E_{\textrm{ext}} = E(2\vec \theta_N - \vec \theta_0)$.
If $E_{\textrm{ext}}\ge E_0$ then moving along the new direction is obviously fruitless.
Another condition is to check if the following condition holds
\begin{equation}
\label{eq:powell_condition}
    2(E_0-2E_N+E_{\textrm{ext}})[(E_0-E_N)-\Delta] \ge (E_0 - E_{\textrm{ext}})^2 \Delta
\end{equation}
Roughly speaking, by rejecting the new direction when Eq.~\ref{eq:powell_condition} is satisfied,
the set of directions $\mathcal{V}$ is directed to the limit where all elements are mutually conjugate.
For a more comprehensive understanding of this condition, readers are encouraged to refer to the original manuscript by Powell~\cite{powell1964efficient}, which provides a clear and concise derivation.

We finalize this session by summarizing the key points of SOAP. The complete algorithm is presented in Algorithm~\ref{algo:soap}.
The algorithm begins by initializing the set of directions $\mathcal{V}$ with unit vectors in the parameter space.
It then performs a line search along each direction in $\mathcal{V}$ to find the minimum energy point. 
The line search involves fitting a parabola based on 2 to 4 data points and using it to estimate the minimum.
The line search procedure is specifically designed for UCC ansatz and is significantly more efficient and noise-resilient than general line search procedures such as the golden section search.
One advantage of SOAP is its ability to handle cases where the circuit parameters are correlated, which is a limitation of traditional sequential optimization methods. After an iteration over $\mathcal{V}$, SOAP identifies promising average directions that account for the correlation between parameters and updates $\mathcal{V}$ with the new direction.
The algorithm then proceeds to the next iteration until certain convergence criteria are satisfied.

\begin{algorithm}
\DontPrintSemicolon
\caption{Sequential Optimization with an Approximate Parabola (SOAP)}
\label{algo:soap}
\SetKw{Break}{break}
\SetKw{Continue}{continue}
\SetKwFunction{LSAP}{LSAP}
\SetKwFunction{UpdateDirection}{UpdateDirection}
\SetKwFunction{MeasureEnergy}{MeasureEnergy}
\SetKwInOut{Input}{input}\SetKwInOut{Output}{output}

\Input{MP2 amplitudes}
\Output{Circuit parameter at the minimum $\vec \theta^*$ and the corresponding energy $E(\vec \theta^*)$}
\BlankLine
Set $\vec \theta_N$ to MP2 amplitudes \;
$E_N = $ \MeasureEnergy{$\vec \theta_N$} \;
Initialize $\mathcal{V}$ by $\vec v_i = \vec e_i$ \;
\While{not converged}{
    $\vec \theta_0 = \vec \theta_N, E_0 = E_N$ \;
    \For{each $\vec v_i$ in $\mathcal{V}$}{
        $\vec \theta_i$, $E_i$ =  \LSAP{$\vec v_i$, $\vec \theta_{i-1}$, $E_{i-1}$} \tcc*{See Algorithm~\ref{algo:line_search}}
        $\Delta_i = E_{i-1} - E_i$ \;
        \If{$\Delta_i$ is the largest so far in the iteration}{$j=i, \Delta=\Delta_i$}
    }
    \tcc{Update $\mathcal{V}$ using the Powell's algorithm}
    $E_\textrm{ext}$ = \MeasureEnergy{$2\vec \theta_N - \vec \theta_0$} \;
    \lIf{$E_{\rm{ext}} \ge E_0$}{\Continue}
    \tcc{Eq.~\ref{eq:powell_condition} is an inequality of $E_0$, $E_N$, $E_{\rm{ext}}$ and $\Delta$}
    \lIf{Eq.~\ref{eq:powell_condition} is satisfied}{\Continue } 
    Remove $\vec v_j$ from $\mathcal{V}$ \;
    Add $(\vec \theta_N - \vec \theta_0) / |\vec \theta_N - \vec \theta_0|$ to the beginning of $\mathcal{V}$ \;
}
\end{algorithm}

\subsection{Numerical Details}
\label{sec:num_detail}
For benchmark results we use 5 different molecular systems in this paper.
For UCC ansatz, the molecules are \ch{N2}, \ch{H8} chain at even spacing and tetrahedral \ch{CH4}.
The bond length $d$ for \ch{H8} and \ch{CH4} is defined as the distance between H atoms and the distance between the C and H atoms, respectively.
The STO-3G basis set is used throughout with
the $1s$ orbital for \ch{N2} and \ch{CH4} frozen.
Under the Jordan-Wigner transformation, all 3 systems correspond to 16 qubits.
We focus on the UCC with singles and doubles (UCCSD) ansatz as a representative UCC ansatz.
\rev{For HEA ansatz, we employ \ch{LiH} and \ch{BeH2} with STO-3G basis,
which corresponds to 10 and 12 qubits respectively with parity transformation.}
The numerical simulation of quantum circuits and hardware experiments are conducted via TensorCircuit~\cite{zhang2023tensorcircuit} and TenCirChem~\cite{li2023tencirchem}.
\rev{TenCirChem is equipped with an efficient algorithm to simulate UCC circuits that allows 16-qubit simulation in minutes on a laptop.
Additionally, TenCirChem evaluates the gradients by auto-differentiation with reversible computing which enables the computation of gradients with $O(1)$ time and space complexity.}
The \textit{ab initio} integrals, reference energies, and MP2 amplitudes are calculated via the PySCF package~\cite{sun2018pyscf, sun2020recent}.

\section{Results and Discussion}
\label{sec:results}

\subsection{The energy landscape}
\begin{figure}[h]
    \centering    \includegraphics[width=\textwidth]{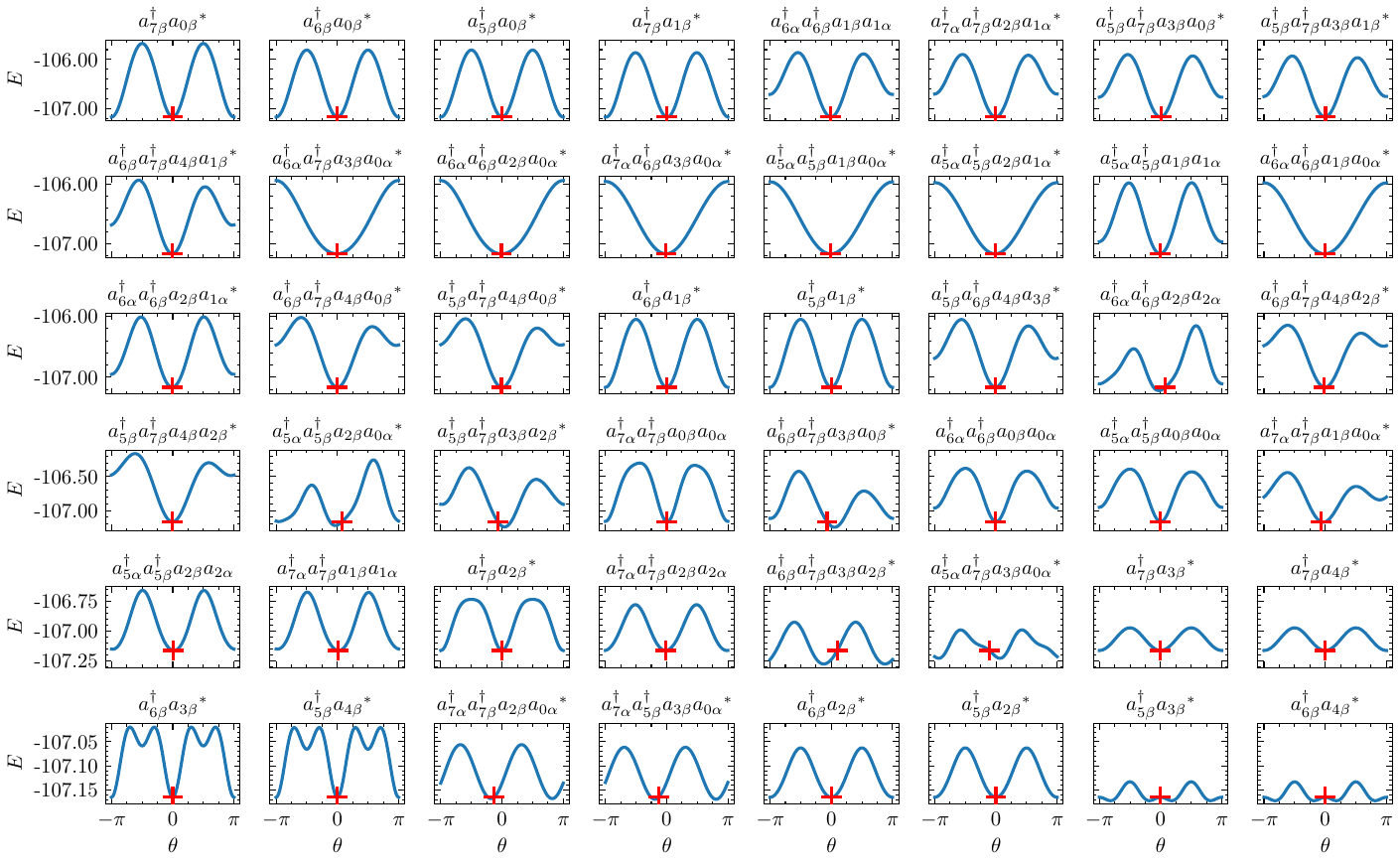}
    \caption{Energy landscape with respect to a single parameter for the UCC ansatz at the initial point generated by MP2. \ch{N2} molecule with bond length 2.0 Å is used as the model system. The energy unit is Hartree. The red crossing represents the MP2-predicted amplitudes and the corresponding energy. The excitation operators associated with each parameter are shown in the title. The Hermitian conjugation is omitted for brevity.
    \rev{If an asterisk is shown, it represents two excitation operators sharing the same parameter.}}
    \label{fig:landscape_n2}
\end{figure}

The success of SOAP relies on the assumption that the landscape can be approximated by a quadratic function, i.e., a parabola.
The assumption is justified by theoretical analysis in Sec.~\ref{sec:rev-mot}.
In this section, we provide numerical evidence to further validate this assumption. 
The details of the model systems and computational methods are presented in Sec.~\ref{sec:num_detail}.
We examine the landscape of the energy function with respect to a single parameter while keeping the other parameters fixed at the initial guess generated by MP2.
We use the $\ch{N2}$ molecule with a bond length of $d=2.0$ Å as an example system.
The results of the landscape analysis are shown in Fig.~\ref{fig:landscape_n2}.
At the top of each sub-figure, the excitation operators associated with each parameter are displayed.
The corresponding Hermitian conjugation is omitted for brevity.
\rev{
If an asterisk is shown, it represents two excitation operators sharing the same parameter}.
The red crossing represents the amplitude of the initial guess and the corresponding energy.
From Fig.~\ref{fig:landscape_n2} it is evident that in most cases the MP2 initial guess is very close to the minimum energy point.
As a consequence, a quadratic expansion around the initial guess appears to be a  reasonable approximation.
Meanwhile, if the starting point deviates from the minimum, a simple update as described in  Algorithm~\ref{algo:line_search} should be sufficient to guide the parameter to the minimum.
The $\ch{N2}$ molecule at stretched bond length is considered a prototypical example of strongly correlated systems.
Therefore, it is expected that for less correlated systems the parabola expansion assumption holds even more valid.
\rev{More examples of the landscape, including \ch{N2} but with a zoomed-in $x$ axis and the landscape of \ch{CH4}, are shown in the Supporting Information.}

\begin{figure}[h]
    \centering    \includegraphics[width=0.6\textwidth]{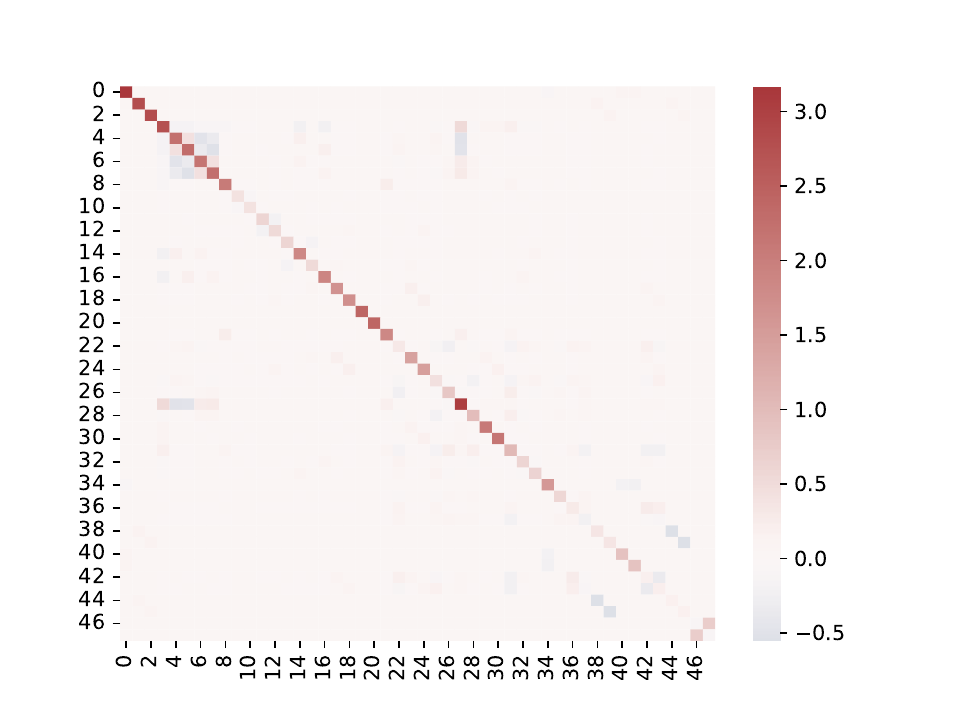}
    \caption{The Hessian matrix for the UCC ansatz at the initial parameter generated by MP2. \ch{N2} molecule with bond length 2.0 Å is used as the model system. The order of the parameters is the same as Fig.~\ref{fig:landscape_n2}.}
    \label{fig:hessian}
\end{figure}

In order to further investigate the landscape of the energy function and at the proximity to the minimum, we examine the Hessian matrix corresponding to the UCC ansatz shown in Fig.~\ref{fig:landscape_n2}.
The Hessian matrix is estimated numerically using the numdifftools package~\cite{BrodtkorbAndDErrico2015}.
In Fig.~\ref{fig:hessian}, we present a plot of the Hessian matrix. The elements with the largest amplitudes lay in the diagonal of the Hessian matrix, which means that most of the parameters in the ansatz exhibit large second-order derivatives. 
Meanwhile, the contribution of non-diagonal elements in the Hessian matrix is non-negligible.
These non-diagonal elements imply that corresponding excitations could lead to a significant energy decrease if they are excited simultaneously.
This indicates the presence of correlations between the parameters and emphasizes the need for optimization methods that account for such correlations.

\subsection{Performance of SOAP}
\begin{figure}[H]
    \centering    \includegraphics[width=\textwidth]{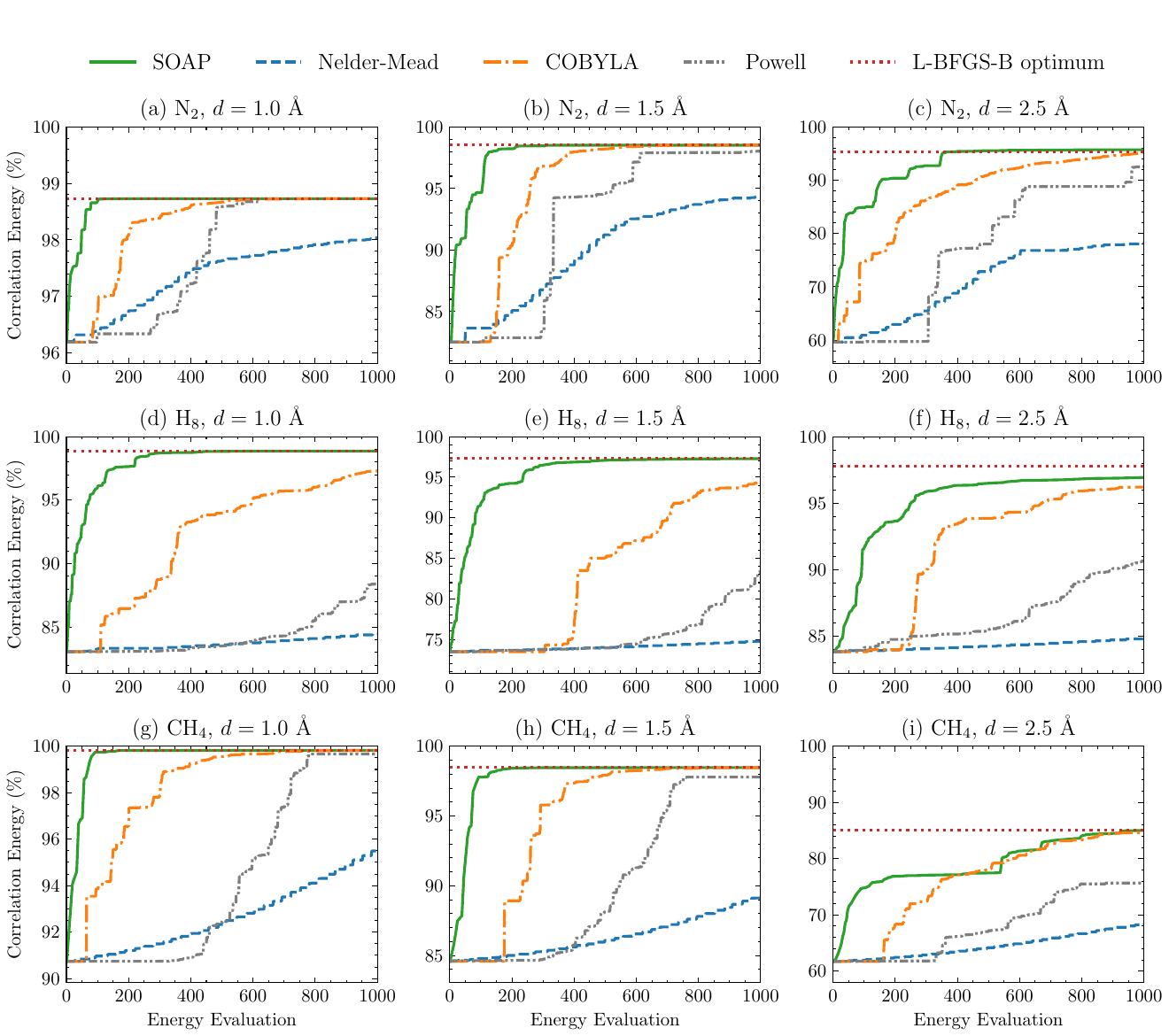}
    \caption{The convergence rate of SOAP compared to other gradient-free optimization methods based on noiseless simulation.  The \rev{red} dotted line is the optimum value from the L-BFGS-B method, which is a gradient-based method that is costly on a quantum device.  The (a)-(c), (d)-(f), and (g)-(i) panels are for the \ch{N2}, \ch{H8} and \ch{CH4} molecules respectively.
    From left to right the bond length increases from \rev{$d=1.0$} Å to $d= 2.5$ Å.}
    \label{fig:noiseless}
\end{figure}

To evaluate the performance of SOAP \rev{on the UCCSD ansatz}, we conducted experiments on three model systems: \ch{N2}, \ch{H8}, and \ch{CH4}, as described in Section~\ref{sec:num_detail}. 
The number of independent parameters for these systems is 48, 108, and 62, respectively.
In our comparison, we included three traditional gradient-free optimization algorithms: Nelder-Mead algorithm~\cite{nelder1965simplex}, COBYLA, and Powell's method. 
We intentionally excluded gradient-based methods from our comparison
due to their higher quantum resource demand.
These methods typically require more energy evaluations than gradient-free methods, because calculating energy gradients via the parameter-shift rule is very expensive. 
More specifically, for the three model systems examined in this work, the number of energy evaluations required to compute the gradients \rev{using parameter-shift rule} is 936, 2816, and 1344, respectively.
\rev{The number is much higher than the number of independent parameters because the parameter-shift rule requires each variable to be associated with a single-qubit rotation gate.}
As presented in Fig.~\ref{fig:noiseless} and Table~\ref{tab:steps}, SOAP can reach convergence using fewer energy evaluations than what is needed to compute the gradients even once.
Additionally, the SPSA algorithm is not included either, as it reduces to a finite-difference-based stochastic gradient descent method when the initial guess is close to the minimum. 
\rev{
In our test calculations, we have observed that SPSA performs poorly for the UCC ansatz, particularly when measurement noise is present. 
This finding contradicts a recent paper where SPSA is reported to outperform COBYLA~\cite{bonet2023performance}.
We conjecture that the reason for this discrepancy is that our benchmark system is significantly larger, and we did not conduct extensive hyper-parameter tuning.
}

Fig.~\ref{fig:noiseless} shows the optimized energy versus the number of energy evaluations by the optimizer using a noiseless simulator.
The energy values are rescaled by the correlation energy $E_{\textrm{corr}} = E_{\textrm{HF}} - E_{\textrm{FCI}}$,
whose exact values are listed in Table~\ref{tab:e_corr}.
The optimal energy obtained by the L-BFGS-B method is used as the best possible energy for the UCC ansatz.
It is worth noting that although the L-BFGS-B method performs well for numerical simulations where the energy gradients are available in $\mathcal{O}(1)$ time and space complexity,
it requires $\mathcal{O}(N)$ energy evaluations when executed on real quantum devices.
\rev{Besides, L-BFGS-B also degrades when measurement noise is present.}
The results \rev{in Fig.~\ref{fig:noiseless}} demonstrate that SOAP exhibits significantly faster convergence than traditional optimization methods, particularly at intermediate bond lengths. 
As expected, at the dissociation limit where $d = 2.5$ Å, the performance gap between SOAP and other methods becomes smaller because the initial guess runs away from the minimum due to strong correlation effects.
Nonetheless, even in these cases, SOAP still demonstrates the fastest convergence except for \ch{CH4} at $d=2.5$ Å.
\rev{
In the Supporting Information, we include benchmark results where the initial parameter is set to the HF state or random state.
And from the results the relative performance between SOAP and traditional optimizers is not sensitive to the initial parameters.
} 
Among the traditional optimization methods, the COBYLA method appears to be the most efficient, which is consistent with previous studies~\cite{singh2023benchmarking}.
\rev{We also note that the MP2 initialization appears to significantly accelerate convergence. For equilibrium bond distance, the MP2 initialization allows the optimizers to start the optimization from approximately 90\% correlation energy, compared with 0\% correlation energy with HF initialization.
More data with $d= 0.5$ Å and $d= 2.0$ Å is provided in the Supporting Information.}

\newcolumntype{s}{>{\hsize=.5\hsize}X}
\begin{table}[]
    \caption{$E_{\textrm{corr}}$ for \ch{N2}, \ch{H8} and \ch{CH4} from $d=0.5$ Å to $d= 2.5$ Å. The unit is Hartree.}
    \label{tab:e_corr}
\begin{tabularx}{0.8\textwidth}{Xsssss}
\toprule
       &  \multicolumn{5}{c}{Bond length}   \\
{Molecule} &   0.5 Å &     1.0 Å &     1.5 Å &     2.0 Å &     2.5 Å \\
\hline
\ch{N2}  &  0.0374 &  0.1294 &  0.3090 &  0.5836 &  0.8234 \\ 
\ch{H8}  &  0.0529 &  0.1332 &  0.3234 &  0.6353 &  0.9208 \\ 
\ch{CH4} &  0.0277 &  0.0660 &  0.1698 &  0.3678 &  0.6238 \\ 
\hline
\end{tabularx}
\end{table}

To provide a quantitative assessment of the relative performance of the optimization methods, we compare the number of energy evaluations required for convergence in Table~\ref{tab:steps}.
Here the convergence is defined as reaching 99\% of the correlation energy obtained by the L-BFGS-B method.
For example, let's consider a scenario where the correlation energy is 0.150 Hartree and the optimal correlation energy obtained by the L-BFGS-B method is 0.100 Hartree.
In this case, convergence is considered as reached if the optimizer reaches a correlation energy of $0.100\times99\%=0.099$ Hartree.
While this criterion cannot be practically used to determine convergence since the energy by the L-BFGS-B method is not known in advance, it serves as a fair measure of how quickly the optimizer finds the minimum defined by the UCC ansatz.
The results presented in Table~\ref{tab:steps} demonstrate that the SOAP optimizer consistently exhibits the fastest convergence among all the methods studied. In many cases, particularly near equilibrium, SOAP is more than five times faster than the second-best method. 
Notably, at a bond length of $d=1.0$ Å, SOAP requires approximately $N$ energy evaluations for an optimization problem with $N$ parameters.
This observation leads us to the conclusion that it is unlikely for any new optimizer to significantly outperform SOAP under such circumstances.

\begin{table}[]

    \caption{The number of energy evaluations required for convergence for SOAP as well as traditional gradient-free optimizers. For a fair comparison, the convergence criterion for all methods is set to reach  99\% of the maximum correlation energy by the UCC ansatz, which is defined by the optimized energy obtained by the L-BFGS-B method. The dashed lines indicate that more than 2000 evaluations are required.}
    \label{tab:steps}
    \centering
    \begin{tabularx}{\textwidth}{>{\hsize=.7\hsize}XXXsssss}
    \toprule
       & & &  \multicolumn{5}{c}{Bond length}   \\
    Molecule &\#Params & Method &   0.5 Å &     1.0 Å &     1.5 Å &     2.0 Å &     2.5 Å        \\
    \hline
    \ch{N2} & 48 & SOAP &      9 &     37 &    116 &    354 &    348 \\
            & & Nelder-Mead &    375 &    630 &  --- &  --- &  --- \\
       & & COBYLA &    236 &    178 &    374 &    606 &    864 \\
       & & Powell &    510 &    441 &    612 &  --- &   1132 \\
    \ch{H8} & 108 & SOAP &    123 &    222 &    286 &    404 &    744 \\
       & & Nelder-Mead &  --- &  --- &  --- &  --- & --- \\
       & & COBYLA &    966 &   1144 &   1312 &   1770 &   1502 \\
       & & Powell &   1205 &   1229 &  --- &  --- &  --- \\
    \ch{CH4} & 62 & SOAP &     39 &     67 &     89 &    144 &    799 \\
       & & Nelder-Mead &   1175 &  --- &  --- &  --- &  --- \\
       & & COBYLA &    304 &    312 &    432 &    562 &    853 \\
       & & Powell &    684 &    733 &    751 &    753 &  --- \\
    \hline
    \end{tabularx}
\end{table}

The most important part of the section is the performance of SOAP in the presence of noise from quantum devices.
The ideal optimizer for parameterized quantum circuits should be noise-resilient, otherwise, the optimizer is only suitable for noiseless numerical simulation.
For the results presented below, we consider the statistical measurement uncertainty, modeled using Gaussian noise.
In traditional numerical simulation of quantum circuits, the measurement uncertainty is modeled by shot-based simulation, which most faithfully emulates the behavior of quantum computers.
However, such a method has a large simulation overhead that limits the simulation to small systems.
To address this issue, we model the measurement uncertainty as a Gaussian noise that is applied to the noiseless energy.
Such a method has negligible computational overhead, and the scale of noisy simulation is thus the same as noiseless simulation.
Gate noise is not taken into consideration for two main reasons.
Firstly, under a moderate noise ratio, gate noise primarily causes a shift in the entire energy landscape, resulting in incorrect absolute energy estimation. However, it does not significantly alter the optimization behavior. 
For a high noise ratio, it is unlikely that VQE will produce meaningful results, and there's no need to be concerned with optimization problems.
Secondly, accurately modeling gate noise requires density matrix simulation or extensive Monte-Carlo simulation, which are considerably more computationally expensive compared to the noiseless case using the statevector simulator.

\begin{figure}[H]
    \centering    \includegraphics[width=\textwidth]{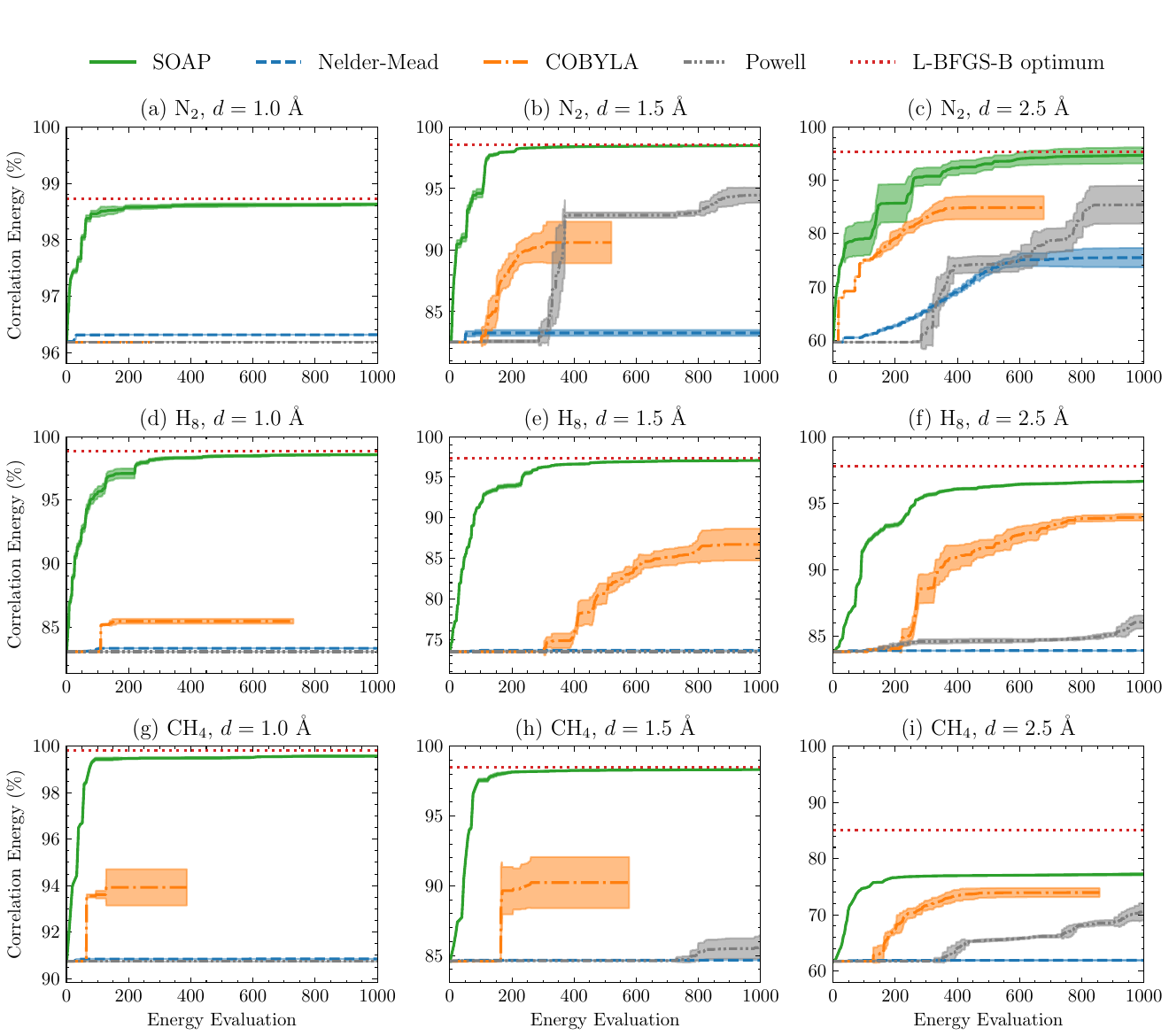}
    \caption{The convergence rate of SOAP compared to other gradient-free optimization methods based on noisy simulation. The quantum noise is modeled as a Gaussian noise to the measured energy with a standard deviation 0.001 mH. The standard deviation from 5 independent trajectories is depicted as the shaded area.
    The \rev{red} dotted line is the optimum value from the L-BFGS-B method based on noiseless simulation.  
    The (a)-(c), (d)-(f), and (g)-(i) panels are for the \ch{N2}, \ch{H8} and \ch{CH4} molecules respectively.
    From left to right the bond length increases from \rev{$d=1.0$} Å to $d= 2.5$ Å. }
    \label{fig:noise}
\end{figure}

In Fig.~\ref{fig:noise} we show the optimizer trajectories when Gaussian noise with a standard deviation of 0.001 Hartree is added to the energy.
For each method, we run 5 independent optimization trajectories, and the standard deviation is depicted as the shaded area.
The results demonstrate that the performance of traditional optimization methods degrades to varying extents when measurement noise is introduced. 
When $d = 1.0$ Å, all of the traditional optimization methods fail catastrophically.
This is because in these cases $E_{\textrm{corr}}$ is typically less than or around 0.1 Hartree (see Table~\ref{tab:e_corr})
and the initial guess has already recovered $\approx 90\%$ of the correlation energy.
Therefore, the presence of 0.001 Hartree noise can easily misdirect the traditional methods, leading to incorrect optimization directions or false convergence.
In contrast, since SOAP is specifically designed to handle measurement uncertainty, it suffers from the least degradation when noise is introduced in the optimization process.
SOAP also exhibits negligible variation over different optimization trajectories.
This resilience to noise can be attributed to the fact that
even when the parameter is close to the minimum, SOAP determines the minimum at each optimization step using a finite step size $u=0.1$.
This approach significantly contributes to SOAP's ability to handle noise and maintain robust optimization performance.
\rev{Similar to Fig.~\ref{fig:noiseless}, we include more data with $d= 0.5$ Å and $d= 2.0$ Å in the Supporting Information.}

\begin{figure}[h]
    \centering    \includegraphics[width=.5\textwidth]{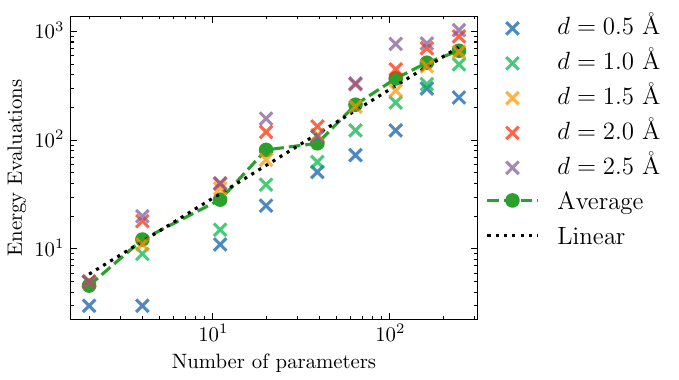}
    \caption{The number of energy evaluations required for convergence for SOAP as a function of the number of parameters in the ansatz.
    The different number of parameters is from hydrogen chains from 2 to 10 atoms, corresponding to a maximum of 20 qubits. Consistent with Table~\ref{tab:steps}, the convergence criteria is set to reach  99\% of the maximum correlation energy, which is defined by the optimized energy from the L-BFGS-B method.  The green dashed line represents the average number of steps across the five different bond distances\rev{, and the black dotted line represents a linear fit of the average number of steps with an expression of $y=2.93x$.}}
    \label{fig:scaling}
\end{figure}

To investigate the scaling behavior of SOAP, we examined how the number of energy evaluations scales with the number of parameters. We select the homogeneous hydrogen chain as the model system and gradually increase the number of atoms from 2 to 10, resulting in the largest system composed of 20 qubits. 
In cases where the number of atoms is odd, a positive charge is added to maintain a closed-shell system.
The number of energy evaluations is defined as the number of energy evaluations to achieve convergence, the same as the values presented in Table~\ref{tab:steps}.
The results are depicted in Fig.~\ref{fig:scaling}, where
five different bond distances from $d=0.5$ Å to $d= 2.5$ Å are considered for each system size.
The green dashed line represents the average number of steps across the five different bond distances.
\rev{
Even in scenarios involving strong correlation, specifically at a bond length of 2.5 Å,  empirically the number of energy evaluations for convergence scales linearly with the number of parameters.
}\rev{
The fitted black line is $y=2.93x$, which means that on average approximately $3N$ energy evaluations are required for convergence if there are $N$ parameters.
This finding is also consistent with the data in Table~\ref{tab:steps}.
}
This favorable scaling can be considered the best-case scenario for a gradient-free optimizer.
Since the number of parameters $N$ scales as $M^4$  for the UCCSD ansatz where $M$ is the system size,
the total scaling, including optimization, is $M^3 \times M^2 \times M^4=M^9$, assuming
an optimistic scaling of circuit depth as $M^3$~\cite{lee2018generalized} and the number of measurements for energy evaluation as $M^2$~\cite{huggins2021efficient}.
\rev{
The $M^3$ scaling for circuit depth is obtained by assuming there are $M^4$ gates in the circuit and $M$ gates can be executed in parallel. 
The $M^2$ measurement scaling is achieved by appending a depth $M$ basis rotation circuit for measurement, allowing $M^2$ terms to be measured simultaneously.
}
To reduce the overall scaling, the key lies in minimizing the number of excitation operators included in the ansatz. This reduction helps decrease both the circuit execution time and the number of energy evaluations required for optimization. 
Despite recent advances~\cite{grimsley2019adaptive, cao2022progress, filip2022reducing, burton2023exact, burton2023accurate}, achieving this reduction while maintaining high accuracy poses a significant challenge.

\begin{figure}[H]
    \centering    \includegraphics[width=0.65\textwidth]{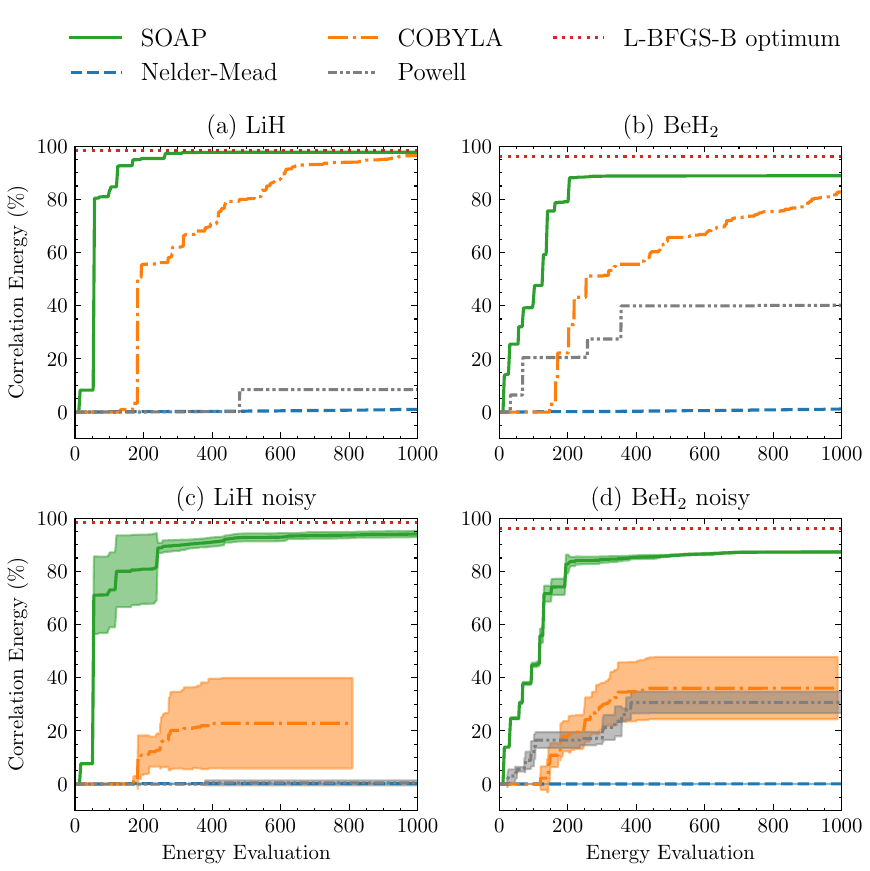}
    \caption{\rev{The convergence rate of SOAP compared to other gradient-free optimization methods based on hardware-efficient ansatz. The results in (a) and (b) are from noiesless simulation and the results in (c) and (d) take measurement uncertainty into account. The quantum noise is modeled as a Gaussian noise to the measured energy with a standard deviation 0.001 mH. The standard deviation from 5 independent trajectories is depicted as the shaded area in (c) and (d). The red dotted line is the optimum value from the L-BFGS-B method based on noiseless simulation.  }}
    \label{fig:chem}
\end{figure}

\rev{
Next, we show that SOAP is a general method that is not limited to the UCCSD ansatz. 
We employ HEA with Clifford-based Hamiltonian Engineering for Molecules (CHEM) that ensures the initial state is HF state and the initial gradients are maximized~\cite{sun2024toward}.
The CHEM algorithm finds a Clifford transformation $\ourmethodU{}$ that transforms the Hamiltonian based on the HEA circuit topology.
For the results presented below we employed a 3-layered HEA with linear staircase CNOTs gates as entanglers.
Each layer contains $R_x$, $R_y$ and $R_z$ parameterized rotation gates for each qubit,
and the total number of parameters is $N=3lN_{\rm{qubit}}$ where $l$ is the number of layers.
The benchmark molecules are \ch{LiH} with bond distance $d=1.6$ Å and \ch{BeH2} with bond distance $d=1.0$ Å.
After applying the parity transformation, the number of qubits for \ch{LiH} is 10, while for \ch{BeH2} it is 12. The corresponding number of parameters for these molecules is 90 and 108, respectively.
The correlation energies obtained for \ch{LiH} and \ch{BeH2} are 0.0205 Hartree and 0.0261 Hartree, respectively. 
The other setups are the same as Fig.~\ref{fig:noiseless} and Fig.~\ref{fig:noise}.
The results in Fig.~\ref{fig:chem} demonstrate that SOAP exhibits a faster convergence speed than other gradient-free methods, 
particularly when considering measurement noise.
Nevertheless, for \ch{BeH2} SOAP appears to trap in a local minimum, and we are actively working to investigate possible solutions to this issue.
}

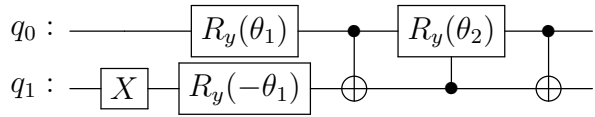
\begin{figure}
    \centering
    \scalebox{1}{
\Qcircuit @C=1.0em @R=0.2em @!R { \\
	 	\nghost{{q}_{0} :  } & \lstick{{q}_{0} :  } & \qw      & \gate{R_y(\theta_1)} & \ctrl{1} & \gate{R_y(\theta_2)} & \ctrl{1} & \qw\\
	 	\nghost{{q}_{1} :  } & \lstick{{q}_{1} :  } & \gate{X} & \gate{R_y(-\theta_1)} & \targ & \ctrl{-1} & \targ & \qw\\
\\ }}
    \caption{The quantum circuit used for hardware experiments.}
    \label{fig:qcircuit}
\end{figure}

\begin{figure}[h]
    \centering    \includegraphics[width=.5\textwidth]{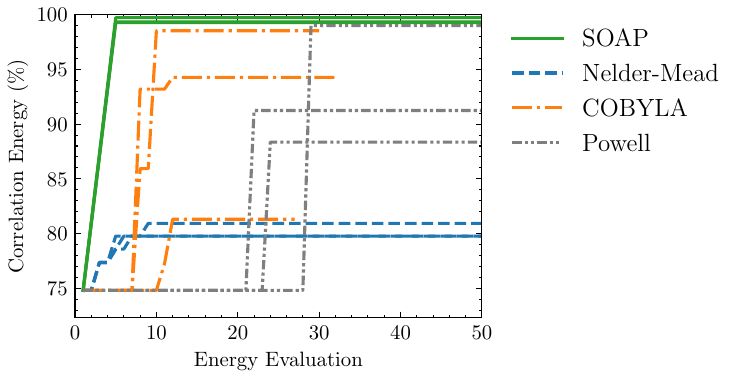}
    \caption{The convergence rate of SOAP compared to other gradient-free optimization methods based on experiments on a superconducting quantum computer.  The benchmark platform is UCCSD calculation of $\ch{H8}$ with a (2e, 2o) active space approximation, which corresponds to 2 qubits and 2 circuit parameters. For each method, 3 independent optimization trials are performed.}
    \label{fig:hea}
\end{figure}

To demonstrate the superiority of SOAP as a parameter optimizer for real quantum devices, we conducted experiments using a superconducting quantum computer. 
\rev{
The superconducting quantum processor is composed of nine transmon qubits and its details are reported in our previous work~\cite{li2023efficient}.
In order to reduce circuit depth}, we use a model system of $\ch{H8}$ and employ a (2e, 2o) active space approximation.
The bond length is set to $d=3.0$ Å, which corresponds to the strong correlation regime.
After parity transformation~\cite{bravyi2002fermionic, seeley2012bravyi}, 2 qubits are required to represent the system wavefunction.
Each term in the Hamiltonian is evaluated using $2^{16}$ measurement shots to determine the expectation value.
\rev{Quantum circuits are susceptible to readout measurement errors, where it is possible to incorrectly read the state  $\ket{0}$ as $\ket{1}$ and vice versa.
To mitigate these errors, standard readout error mitigation techniques were employed in our study.
These techniques aim to correct for the discrepancies between the ideal distribution of bitstrings 
$\vec p_{\rm{ideal}}$ 
  and the noisy distribution of bitstrings 
 $\vec p_{\rm{noisy}}$.
The relationship between the ideal and noisy distributions is described by the response matrix $\Lambda$ such that $\vec p_{\rm{noisy}}=\Lambda \vec p_{\rm{ideal}}$.
The response matrix $\Lambda$ can be decomposed into a direct product of response matrices for each qubit, assuming that the measurements for different qubits are independent.
To construct $\Lambda$, calibration quantum circuits were used, where each $\ket{0}$ qubit is subjected to an H gate before measurement.}
The \rev{UCCSD} ansatz used in the optimization contains three excitation operators, with one double excitation and two single excitations.
Due to spin symmetry, the two single excitations share the same parameter, resulting in two free parameters in the ansatz.
The overall circuit for the ansatz, before being compiled to native gates on the quantum device, is depicted in Fig.~\ref{fig:qcircuit}.
The $X$ gate creates the initial HF state $\ket{0101}$, which under parity transformation corresponds to $\ket{0110}$ and is reduced to $\ket{01}$ by removing the second qubit and the fourth qubit.
The first two single qubit $R_y$ rotation gates are applied for single excitation, 
while the following CNOT gates and controlled $R_y$ gate are applied for double excitation.
In Fig.~\ref{fig:hea}, we present the convergence behavior of SOAP compared to classical gradient-free optimization methods. To reduce the randomness introduced by measurement uncertainty, the optimization process was independently performed three times for each method. The results clearly demonstrate that SOAP exhibits faster convergence and produces consistent trajectories under realistic hardware conditions.

\section{Conclusion and Outlook}
In conclusion, this work introduces SOAP as an efficient and robust parameter optimization method specifically designed for quantum computational chemistry. SOAP demonstrates superior efficiency by leveraging an approximate parabolic expansion of the energy landscape when the initial guess is close to the minimum. Unlike previous sequential optimization methods, SOAP minimizes the number of energy evaluations by incorporating parameter correlations through the update of the direction set $\mathcal{V}$. The algorithm is straightforward to implement and does not require hyper-parameter tuning, making it user-friendly and well-suited for quantum chemistry applications.

The performance of SOAP is extensively evaluated through numerical simulations and hardware experiments. Numerical simulations involve model systems such as \ch{N2}, \ch{H8}, and \ch{CH4}, which correspond to 16 qubits. 
The results demonstrate that SOAP achieves significantly faster convergence rates, 
as measured by the number of energy evaluations required, 
compared to traditional optimization methods like COBYLA and Powell's method, particularly for bond lengths near equilibrium. 
Surprisingly, at dissociation bond length where strong correlation is present, the performance of SOAP does not degrade significantly.
Furthermore, SOAP exhibits remarkable resilience to measurement noise. \rev{Scaling studies up to 20 qubits reveal that on average SOAP requires $3N$ energy evaluations for convergence where $N$ is the number of parameters in the ansatz, indicating its applicability to larger systems. 
We also showcase that SOAP is a general method applicable to HEA.}
The hardware experiments conducted on a (2e, 2o) active space approximation of \ch{H8} validate SOAP's efficiency and robustness under realistic conditions.

While this paper focuses on the UCCSD ansatz, SOAP is also well-suited for the ADAPT-VQE method~\cite{grimsley2019adaptive}, which has a lower circuit depth and good initial guesses during iteration. Additionally, SOAP is a general method that can be applied to other families of ansatz, provided a suitable initial guess is generated~\cite{zhou2020quantum, ravi2022cafqa}. 
In such cases, SOAP is expected to perform well.
\rev{In addition, the quality of the initial circuit parameters also affects the number of energy evaluations required for convergence.
We expect SOAP can achieve faster convergence with more sophisticated initial guess~\cite{filip2022reducing}.}

\section*{Acknowledgement}
We thank Jonathan Allcock for insightful discussions.
Weitang Li is supported by the Young Elite Scientists Sponsorship Program by CAST, 2023QNRC001.

\section*{Code Availability}
The SOAP algorithm is integrated into the open-source \textsc{TenCirChem} package (\url{https://github.com/tencent-quantum-lab/TenCirChem}).

\section*{Competing interests}
The authors declare no competing interests.

\section*{Supporting Information}
The Supporting Information is available free of charge online,
including more detailed data for the energy landscape, the benchmark for the choice of $u$,
the full data for the performance of SOAP in logarithmic scale,
and test calculations using different parameter initialization strategies.

\bibliography{refs}

\clearpage

\begin{figure}[htp]
    \centering
    \includegraphics[width=0.5\textwidth]{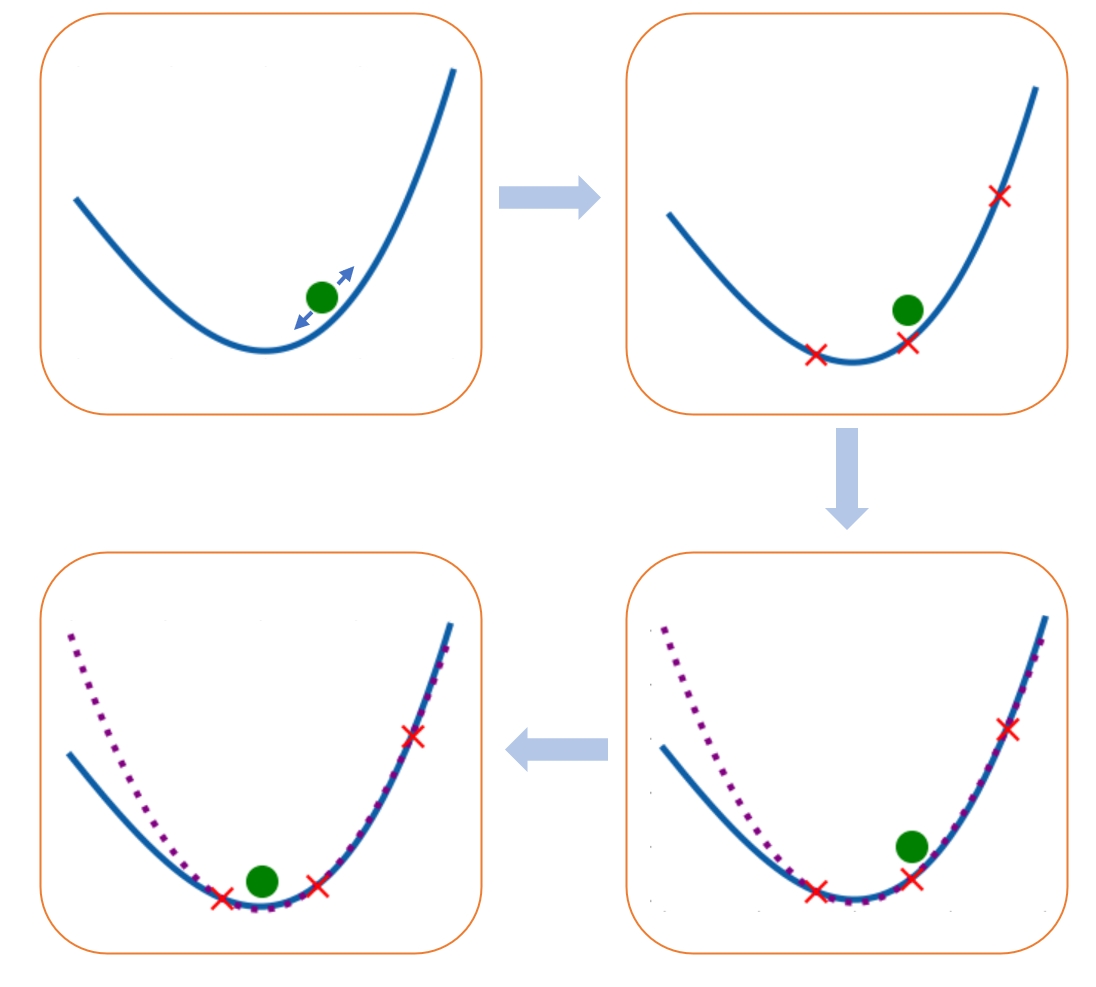}
\caption*{For Table of Contents Only}
\end{figure}

\end{document}


\title{Supporting Information for ``Efficient and Robust Parameter Optimization of the Unitary Coupled-Cluster Ansatz''}

\author{Weitang Li}
\affiliation{Tencent Quantum Lab, Tencent, Shenzhen, China}

\author{Yufei Ge}
\affiliation{Department of Chemistry, Tsinghua University, Beijing, China}

\author{Shi-Xin Zhang}
\affiliation{Tencent Quantum Lab, Tencent, Shenzhen, China}

\author{Yu-Qin Chen}
\affiliation{Tencent Quantum Lab, Tencent, Shenzhen, China}

\author{Shengyu Zhang}
\affiliation{Tencent Quantum Lab, Tencent, Hong Kong, China  \\ \rm{liwt31@gmail.com}}

\maketitle

\section{More Energy Landscapes}

\begin{figure}[H]
    \centering    \includegraphics[width=\textwidth]{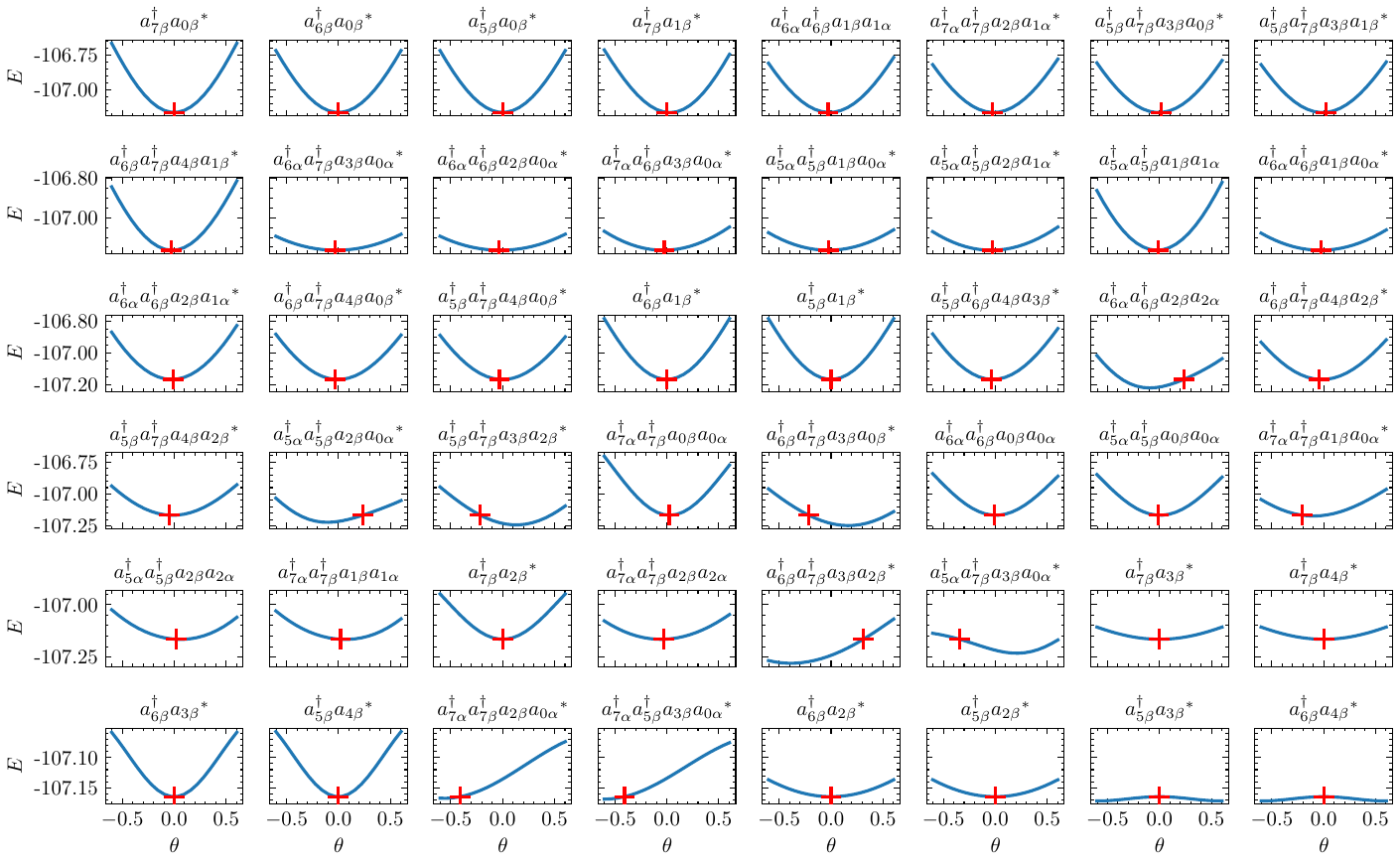}
    \caption{Energy landscape with respect to a single parameter for the UCC ansatz at the initial point generated by MP2. \ch{N2} molecule with bond length 2.0 Å is used as the model system. The energy unit is Hartree. The red crossing represents the MP2-predicted amplitudes and the corresponding energy. The excitation operators associated with each parameter are shown in the title. The Hermitian conjugation is omitted for brevity.
    If an asterisk is shown, it represents two excitation operators sharing the same parameter.}
    \label{fig:landscape_n2_si}
\end{figure}

We first show the energy landscape of \ch{N2} in Fig.~\ref{fig:landscape_n2_si}, identical to Figure 2 in the main text, but with a zoomed-in $x$-axis. In Fig.~\ref{fig:landscape_n2_si} the difference between the MP2 initial guess and the zero point is more visible.

\begin{figure}[H]
    \centering    \includegraphics[width=\textwidth]{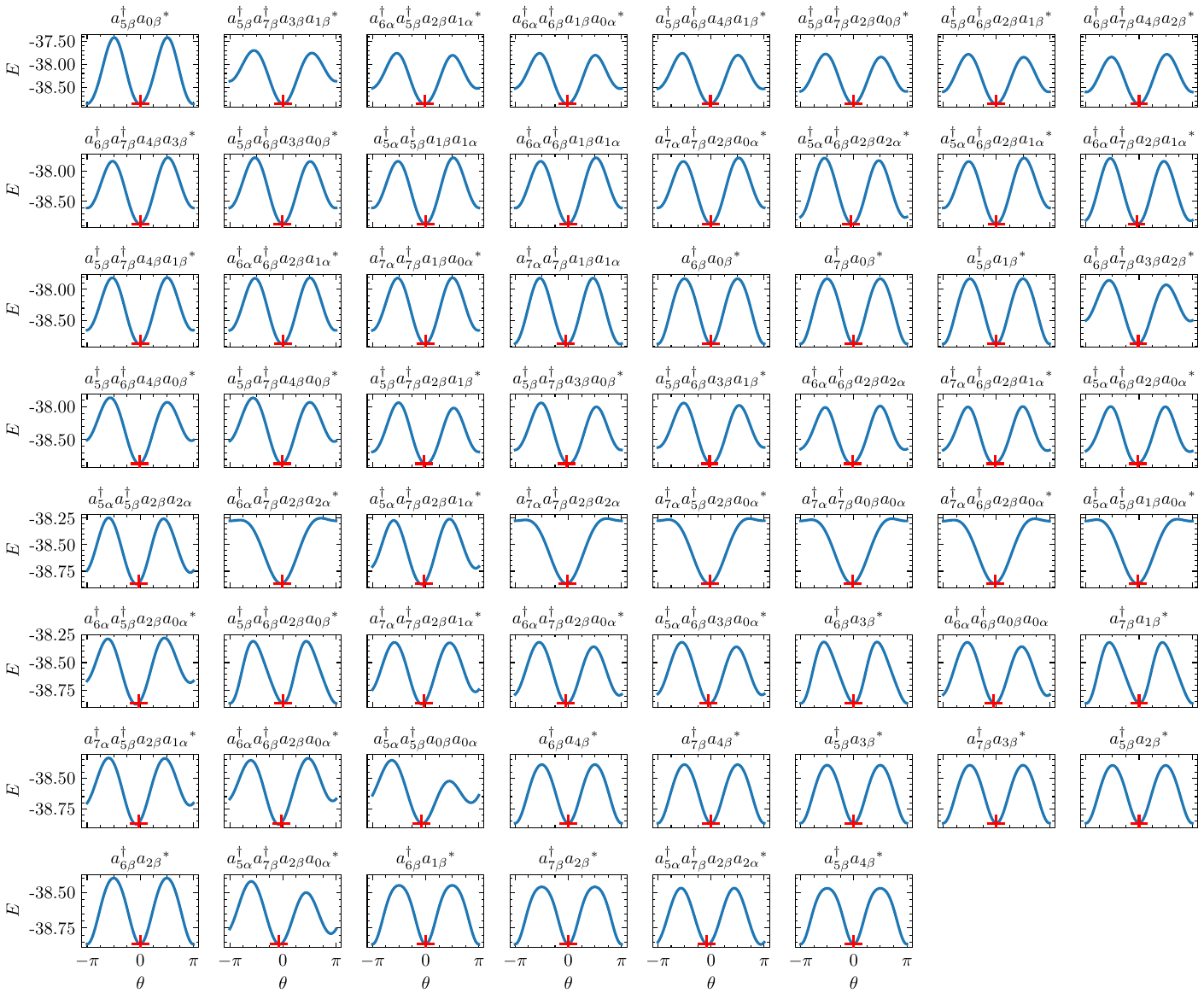}
    \caption{Energy landscape with respect to a single parameter for the UCC ansatz at the initial point generated by MP2. \ch{CH4} molecule with bond length 2.5 Å is used as the model system. The energy unit is Hartree. The red crossing represents the MP2-predicted amplitudes and the corresponding energy. The excitation operators associated with each parameter are shown in the title. The Hermitian conjugation is omitted for brevity.
    If an asterisk is shown, it represents two excitation operators sharing the same parameter.}
    \label{fig:landscape_ch4}
\end{figure}

We next show the energy landscape of \ch{CH4} with bond distance $d=2.5$ Å in Fig.~\ref{fig:landscape_ch4}.
Fig.~\ref{fig:landscape_ch4} shows that for each parameter the initial guess is close to the bottom of a parabola.

\section{Benchmarking Different $u$}
The number of energy evaluations required for convergence for SOAP with different parabola fitting step size $u$ is shown in Table~\ref{tab:steps}.
Noiseless simulator is employed.
While optimal $u$ may vary for specific molecules and bond lengths, overall, the choice of $u$ does not significantly impact the performance.
\newcolumntype{s}{>{\hsize=.5\hsize}X}
\begin{table}[H]
    \caption{The number of energy evaluations required for convergence for SOAP with different $u$. The convergence criterion is set to reach  99\% of the maximum correlation energy by the UCC ansatz, which is defined by the optimized energy obtained by the L-BFGS-B method.}
    \label{tab:steps}
    \centering
    \begin{tabularx}{\textwidth}{>{\hsize=.7\hsize}XXsssss}
    \toprule
       & &  \multicolumn{5}{c}{Bond length}   \\
    Molecule & $u$ for SOAP &   0.5 Å &     1.0 Å &     1.5 Å &     2.0 Å &     2.5 Å        \\
    \hline
    \ch{N2} & $u=0.05$ &      9 &     37 &    140 &    279 &    509 \\
        & $u=0.10$ &      9 &     37 &    116 &    354 &    348 \\
        & $u=0.20$ &    9 &    39 &    120 &    259 &    239 \\
    \ch{H8} & $u=0.05$ &    125 &    230 &    306 &    446 &    822 \\
       & $u=0.10$ &    123 &    222 &    286 &    404 &    744\\
        & $u=0.20$ &    123 &   220 &   268 &   434 &   666 \\
    \ch{CH4} & $u=0.05$ &     39 &     67 &     97 &    178 &    748 \\
        & $u=0.10$ &     39 &     67 &     89 &    144 &    799\\
        & $u=0.20$ &    39 &    71 &    125 &    162 &    664 \\
    \hline
    \end{tabularx}
\end{table}

\section{Optimization Error in Logarithmic Scale}
In this section we provide full data for the optimization performance of SOAP based on noiseless and noisy simulation.
To provide a different view we plot the $y$ axis in logarithmic scale.
It is interesting to note that for the \ch{N2} molecule, when the number of energy evaluations is large, the COBYLA optimizer may eventually outperform the SOAP optimizer. 
This is because SOAP employs a finite $u=0.1$ step size throughout, and its highest accuracy is thus limited.
However, this is a designed feature of SOAP, considering quantum circuit measurement is intrinsically noisy.
In Fig.~\ref{fig:noise_si_log}, we have taken measurement uncertainty into account, resulting in larger errors for the optimizers. In the logarithmic scale, the variation between SOAP and other methods may appear comparable, which is an artifact of the logarithmic scale.

\begin{figure}[H]
    \centering    \includegraphics[width=\textwidth]{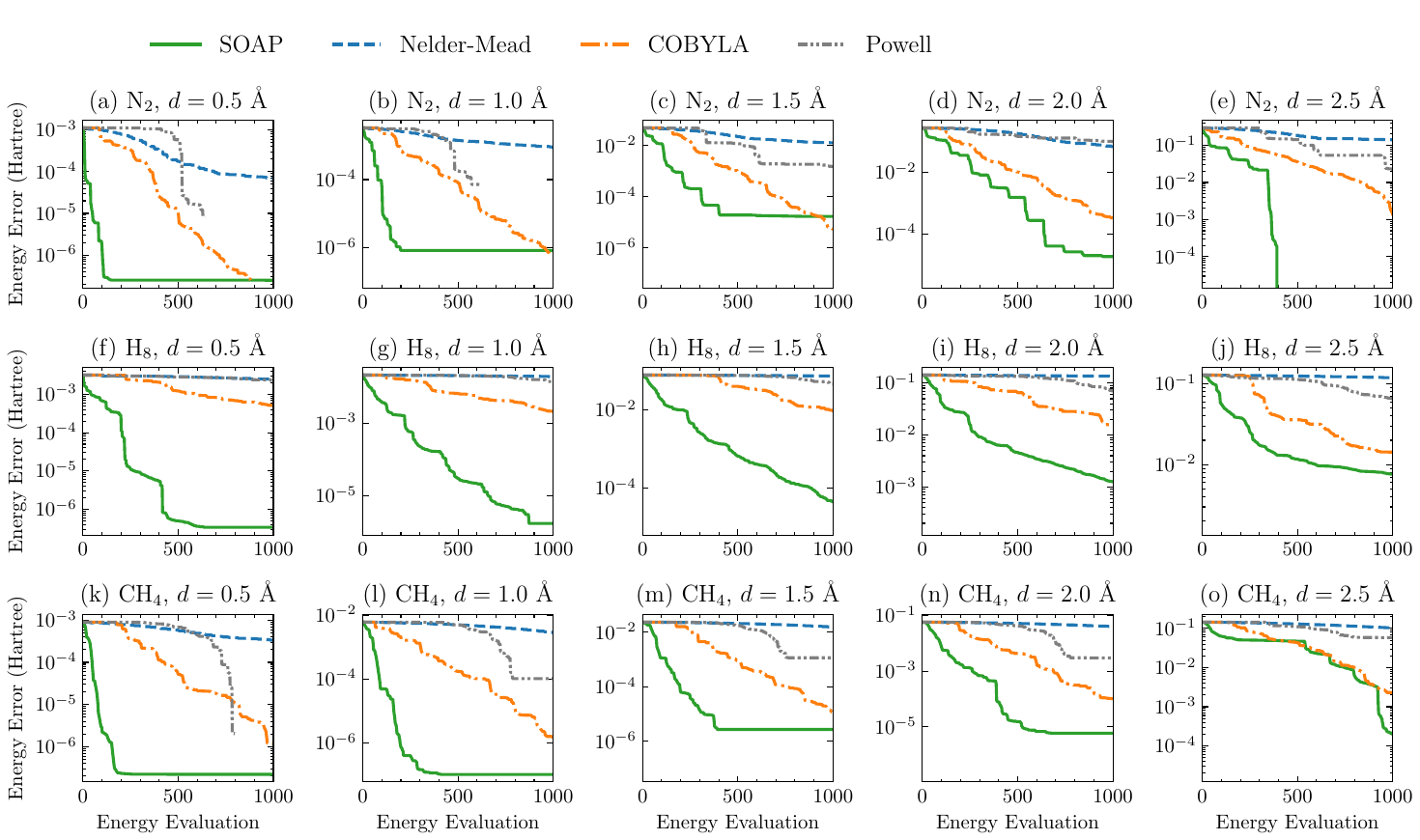}
    \caption{The convergence rate of SOAP compared to other gradient-free optimization methods based on noiseless simulation.
    The reference energy is the optimal energy by the L-BFGS-B optimizer.
    The top, middle, and bottom panels are for the \ch{N2}, \ch{H8} and \ch{CH4} molecules respectively.
    From left to right the bond length increases from $d=0.5$ Å to $d= 2.5$ Å.}
    \label{fig:noiseless_si_log}
\end{figure}

\begin{figure}[H]
    \centering    \includegraphics[width=\textwidth]{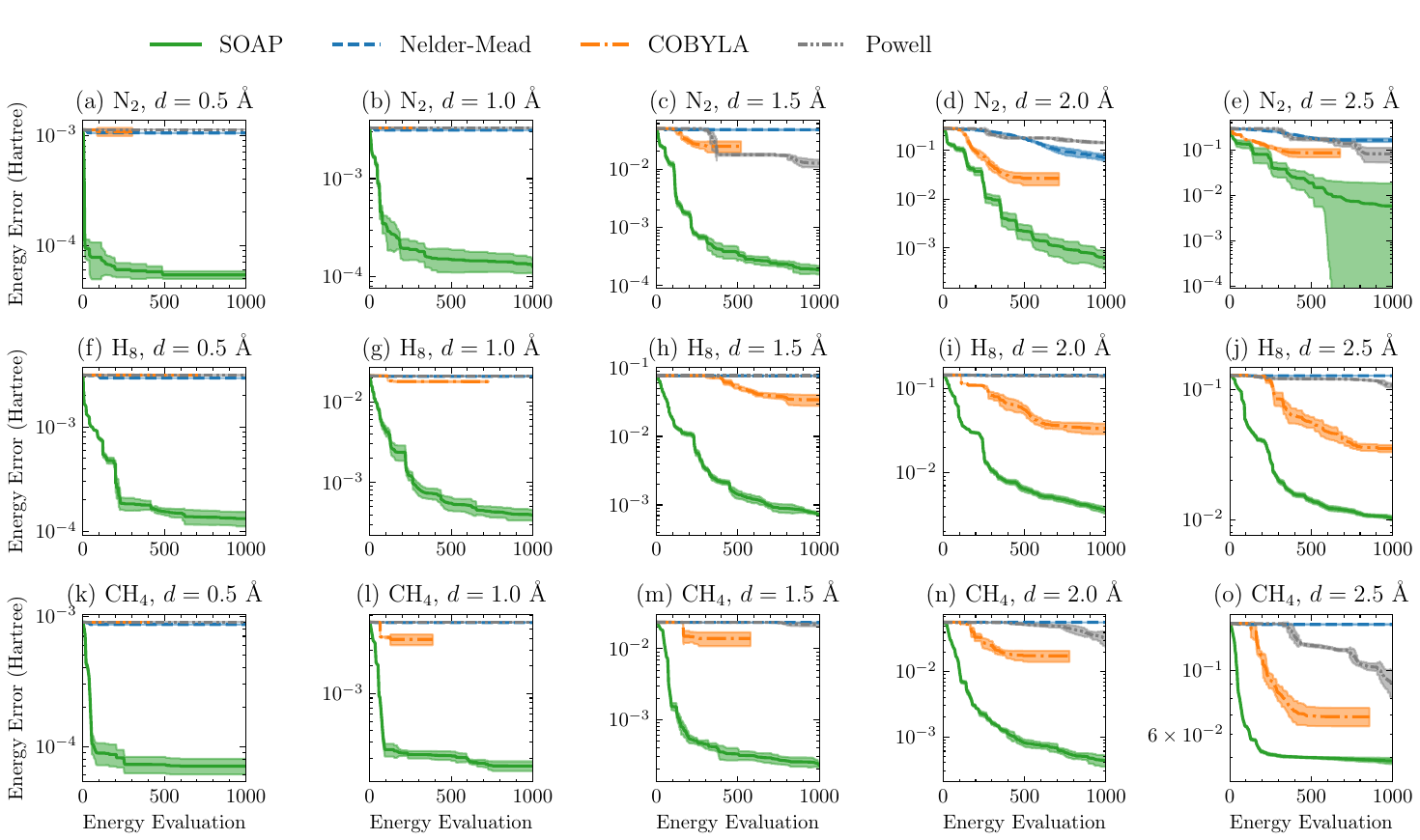}
    \caption{The convergence rate of SOAP compared to other gradient-free optimization methods based on noisy simulation. 
        The reference energy is the optimal energy by the L-BFGS-B optimizer.
    The top, middle, and bottom panels are for the \ch{N2}, \ch{H8} and \ch{CH4} molecules respectively.
    From left to right the bond length increases from $d=0.5$ Å to $d= 2.5$ Å.}
    \label{fig:noise_si_log}
\end{figure}

\section{Effect of Different Initializations}
In this section, we aim to demonstrate the robustness of SOAP, despite being motivated by the fact that the UCC initial guess is close to the minimum. To support this claim, we present data in Table \ref{tab:steps-hf}, which showcases the number of steps required for convergence when using the HF initial state.
By comparing Table~\ref{tab:steps-hf} to Table 2 of the manuscript, we can conclude that when compared to the MP2 initial guess, using the HF initial guess leads to slower convergence for all the optimizers considered. However, SOAP still exhibits the fastest convergence among them.

\begin{table}[]
    \caption{The number of energy evaluations required for convergence for SOAP as well as traditional gradient-free optimizers, with HF initial state. The convergence criterion for all methods is set to reach  99\% of the maximum correlation energy by the UCC ansatz, which is defined by the optimized energy obtained by the L-BFGS-B method. The dashed lines indicate that more than 2000 evaluations are required.}
    \label{tab:steps-hf}
    \centering
    \begin{tabularx}{\textwidth}{>{\hsize=.7\hsize}XXXsssss}
    \toprule
       & & &  \multicolumn{5}{c}{Bond length}   \\
    Molecule &\#Params & Method &   0.5 Å &     1.0 Å &     1.5 Å &     2.0 Å &     2.5 Å        \\
    \hline
    \ch{N2} & 48 & SOAP & 112 & 128 & 231 & 481 & 359  \\
            & & Nelder-Mead &    --- &    --- &  --- &  --- &  --- \\
       & & COBYLA &    557 &    516 &    537 &    --- &    --- \\
       & & Powell &    874 &    962 &    1861 &  --- &   --- \\
    \ch{H8} & 108 & SOAP & 358 & 370 & 489 & 780 & ---\\
       & & Nelder-Mead &  --- &  --- &  --- &  --- & --- \\
       & & COBYLA & --- & 1808 & 1613 & 1920 & 1802 \\
       & & Powell &   1644  &   --- &  --- &  --- &  --- \\
    \ch{CH4} & 62 & SOAP & 128 & 156 & 248 & 423 & 727\\
       & & Nelder-Mead &   --- &  --- &  --- &  --- &  --- \\
       & & COBYLA & 693 & 539 & 625 & 628 & 1101\\
       & & Powell & 737 & 1100 & 1405 & 1486 &   --- \\
    \hline
    \end{tabularx}
\end{table}

We additionally performed test calculations using random initial guesses, and the results are presented below. In these experiments, the initial parameters are uniformly drawn from $[0, 2\pi]$, and we have conducted 5 independent runs with different initial parameter values. These calculations do not consider measurement uncertainty.
Interestingly, despite SOAP not being specifically designed to handle random initial guesses, its performance is comparable and sometimes even better than that of COBYLA, the best traditional optimizer in this task. 

\begin{figure}[H]
    \centering    \includegraphics[width=\textwidth]{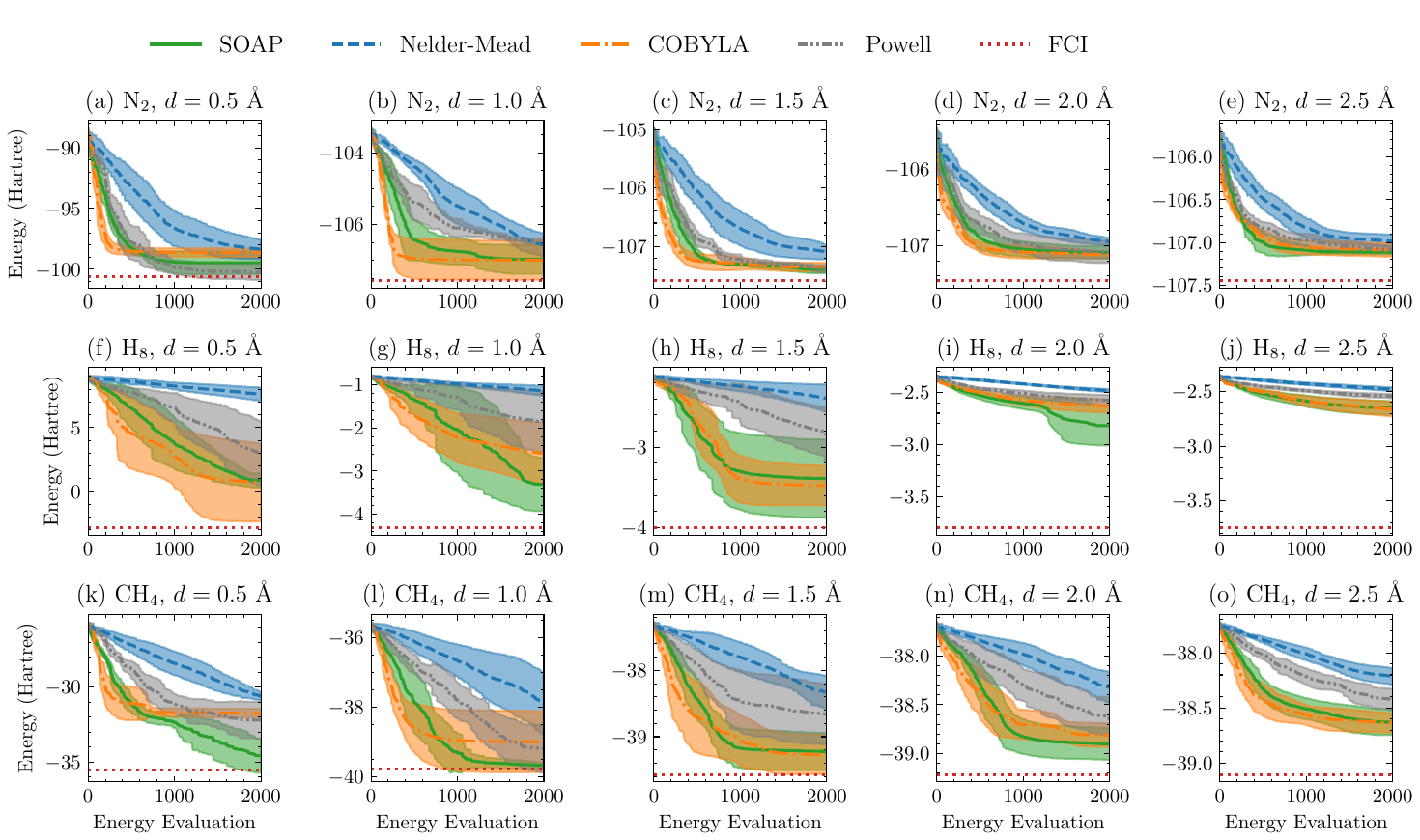}
    \caption{The convergence rate of SOAP compared to other gradient-free optimization methods based on noiseless simulation using random initial guess. The standard deviation from 5 different initial parameters is depicted as the shaded area.
    The top, middle, and bottom panels are for the \ch{N2}, \ch{H8} and \ch{CH4} molecules respectively.
    From left to right the bond length increases from $d=0.5$ Å to $d= 2.5$ Å. }
    \label{fig:noiseless_random}
\end{figure}